\documentclass[12pt,draftcls,onecolumn,twoside]{IEEEtran}
\usepackage{graphicx,amssymb,amsmath,color}
\usepackage{subfigure}
\usepackage{algorithm,theorem}
\usepackage{url}
\usepackage{CJK}
\usepackage{algorithmic,algorithm}
\usepackage[noadjust]{cite}
\usepackage{cases}
\usepackage{enumerate}
\usepackage[square, comma, sort&compress, numbers]{natbib}

\newcommand\Wc{\ensuremath{\mathcal{W}}}
\newcommand\Fc{\ensuremath{\mathcal{F}}}

\newcommand\Xc{\ensuremath{{\mathcal{X}}}}

\newtheorem{Proposition}{Proposition}
\newtheorem{Theorem}{Theorem}

\newtheorem{Remark}{Remark}
\newtheorem{Assumption}{Assumption}

\definecolor{orange}{RGB}{255,107,0}
\definecolor{green} {RGB}{0,180,80}

\title{{Energy-Efficient Packet Scheduling with Finite Blocklength Codes: Convexity Analysis and Efficient Algorithms}}

\author{Shengfeng Xu,
        Tsung-Hui Chang,
        Shih-Chun Lin,
        Chao Shen,
        and~Gang Zhu

\thanks{Part of this work has been presented in IEEE Global Communications Conference (GLOBECOM), San Diego, Dec. 2015 \cite{Xu2015}. This work was supported by the Fundamental Research Funds for the Central Universities (No. 2015YJS032), the State Key Laboratory of Rail Traffic Control and Safety (No. RCS2015ZT001, No. RCS2014ZT34), the Natural Science Foundation of China (No. U1334202 and No. 61571385), the MOST Taiwan (R.O.C.) MOST-104-2628-E-011-008-MY3, and ``Aiming For the Top University Program", Ministry of Education, Taiwan. T.-H. Chang, S.-C. Lin and C. Shen are corresponding authors.}

\thanks{Shengfeng Xu, Chao Shen, and Gang Zhu are with the State Key Laboratory of Rail Traffic Control and Safety, Beijing Jiaotong University, Beijing, China (emails: xsf1988@bjtu.edu.cn; shenchao@bjtu.edu.cn; gzhu@bjtu.edu.cn).
Tsung-Hui Chang is with the School of Science and Engineering, Chinese University of Hong Kong, Shenzhen (email: tsunghui.chang@ieee.org).
Shih-Chun Lin is with the Department of Electronic and Computer Engineering, National Taiwan University of Science and Technology, Taipei, Taiwan (email: sclin@mail.ntust.edu.tw) .
}
}

\begin{document}

{\linespread{1.2} \rm
\maketitle
\vspace{-0.0cm}
\begin{abstract}
This paper considers an energy-efficient packet scheduling problem over quasi-static block fading channels. The goal is to minimize the total energy for transmitting a sequence of data packets under the first-in-first-out rule and strict delay constraints. Conventionally, such design problem is studied under the assumption that the packet transmission rate can be characterized by the classical Shannon capacity formula, which, however, may provide inaccurate energy consumption estimation, especially when the code blocklength is finite. In this paper, we formulate a new energy-efficient packet scheduling problem by adopting a recently developed channel capacity formula for finite blocklength codes. The newly formulated problem is fundamentally more challenging to solve than the traditional one because the transmission energy function under the new channel capacity formula neither can be expressed in closed form nor possesses desirable monotonicity and convexity in general. We analyze conditions on the code blocklength for which the transmission energy function is monotonic and convex. Based on these properties, we develop efficient offline packet scheduling algorithms as well as a rolling-window based online algorithm for real-time packet scheduling. Simulation results demonstrate not only the efficacy of the proposed algorithms but also the fact that the traditional design using the Shannon capacity formula can considerably underestimate the transmission energy for reliable communications.
\end{abstract}
}
\newpage
\section{Introduction}
One of the most urgent tasks in constructing the future 5G communication network is to mitigate the energy
consumption despite greatly increased demands for data transmission rate
\cite{Andrews-2014}. This calls for advanced packet scheduling
designs that not only account for the quality of services (e.g.,
delay constraints and packet error probabilities) but also
minimize the transmission energy expenditure for green communications. Energy-efficient packet scheduling problems have been extensively studied in the literature; see, e.g.,
\cite{Biyikoglu-2002, Chen-2007, Chen-2009, Zafer-2009, Wang-2013}. In particular, work \cite{Biyikoglu-2002} has studied the energy-efficient packet scheduling problem in additive white Gaussian noise (AWGN) channels by
assuming that all packets have a common deadline. Work \cite{Chen-2007} extends \cite{Biyikoglu-2002} to the setting where the packets have individual delay constraints and follow the first-in-first-out
(FIFO) rule.
While \cite{Biyikoglu-2002, Chen-2007} have focused on optimizing the optimal packet blocklength, works
\cite{Zafer-2009,Wang-2013,Chen-2009} aimed at finding the optimal packet transmission rates for minimizing the total transmission energy. Notably, works \cite{Zafer-2009,Wang-2013,Chen-2009} have considered quasi-static block fading channels instead of AWGN channels.

An assumption that is commonly made in the aforementioned works \cite{Biyikoglu-2002, Chen-2007, Chen-2009,
Zafer-2009, Wang-2013} is that the transmission power and rate obey the classical Shannon capacity formula, which, however, is valid only when the channel code has an extremely long length \cite{Polyanskiy-2010}. Unfortunately, the future 5G system is expected to support a wide range of services for emerging
applications, such as metering and traffic safety
\cite{Popovski-2014}, where the packet blocklength is short and sometimes under stringent delay and
reliability constraints \cite{Durisi-2015}. This implies that the
long code required by the Shannon capacity formula may
become prohibitive, and thus
previous designs in \cite{Biyikoglu-2002, Chen-2007, Chen-2009,
Zafer-2009, Wang-2013} do not fit these 5G applications. Therefore, it is
paramount to investigate energy-efficient packet scheduling
problems for finite blocklength codes. In fact,
recent studies in information theory \cite{Polyanskiy-2010} have
also revealed the fact that the Shannon capacity may yield inaccurate
engineering insights into the system design once the
code blocklength is constrained. In view of this, a new channel
capacity formula for the finite blocklength codes has been
developed in \cite{Polyanskiy-2010}, which predicts the rate
performance of short packet transmission more accurately than the
Shannon capacity formula \cite{Yang-2013}.
Note that this new capacity
formula has recently been considered  in
\cite{Gursoy-2013, Ozcan-2013, Makki-2014, Makki-2014WCL, Makki-2015, Hu-2015CL, Hu-2015} for network performance analysis, though
the results therein cannot be  used for packet scheduling designs.

In this paper, we study the energy-efficient packet scheduling
problem using the new capacity formula for the finite blocklength codes
in \cite{Polyanskiy-2010}. Specifically, we consider a scenario where a scheduler wants to transmit a sequence of packets using a minimal transmission
energy subject to FIFO and strict delay constraints over quasi-static block fading channels.
In \cite{Biyikoglu-2002, Chen-2007, Chen-2009, Zafer-2009, Wang-2013}, the packet scheduling algorithms are developed under the assumption that the transmission energy is a \textit{monotonic and convex} function of the blocklength (or rates in \cite{Chen-2009,
Zafer-2009, Wang-2013}). Such desirable properties naturally hold true when the traditional Shannon
capacity formula is adopted. However,
when the new capacity formula for the finite blocklength codes is considered, the packet scheduling problem is
\textit{fundamentally} more challenging. First, the monotonicity and convexity of
the transmission energy function may \textit{no longer} hold in
the finite blocklength regime. Second, due to the complex structure of the new channel capacity formula, the transmission energy function does \textit{not} even have an explicit expression. This implies that the packet scheduling problem cannot be directly solved by off-the-self solvers even if it is a convex optimization problem.
To the best of our knowledge, the current paper
is the first to investigate the energy-efficient packet
scheduling problem for finite blocklength codes. The main
contributions of this paper are summarized as follows.
\begin{itemize}
  \item We formulate the energy-efficient packet scheduling problem for finite blocklength codes over quasi-static block fading channels. Specifically, we consider the finite-blocklength channel capacity formula in \cite{Polyanskiy-2010} and propose to jointly optimize
      the packet transmission power and code blocklength to minimize the total transmission energy subject to strict delay constraints.
      By applying the implicit function theorem \cite{Krantz_Parks02}, we analytically show that the energy function under the finite blocklength codes can still preserve the monotonicity and convexity under mild conditions on the code blocklength.

  \item Two offline packet scheduling algorithms are proposed to solve the considered packet scheduling problem. When the packet scheduling problem is a (strictly) convex problem, we show that the multi-level water filling (MLWF) algorithm in \cite{Wang-2013} can be modified to solve the considered packet scheduling problem. In particular, the analyzed monotonicity and convexity of the energy function can be utilized to implement the MLWF algorithm without the need of knowing explicit expression of the energy function. For the general case where the objective function may not be convex, we modify the successive upper-bound minimization (SUM) method in \cite{Razaviyayn-2013BSUM}
      to handle the considered packet scheduling problem in the absence of explicit expression of the energy function.
      Inspired by the proposed offline scheduling algorithms, we further develop an online algorithm for real-time packet scheduling.
  \item Numerical simulations are conducted to demonstrate the importance of the newly formulated packet scheduling problem and the performance advantage of the proposed algorithms. Simulation results show that the transmission energy required by the finite-blocklength packet scheduling problem is higher than that using the traditional Shannon capacity formula, which illustrates the fact that the latter may considerately underestimate the required transmission energy for reliable communications.
      Specifically, based on our numerical results, the traditional design underestimates about 10$\%$ transmission energy for achieving $5\times10^{-4}$ packet error probability.
\end{itemize}

{\bf Synopsis:}  Section \ref{sec: sys model} describes the system model and the finite blocklength packet scheduling problem. In Section \ref{sec: mono and cvxit}, the monotonicity and convexity properties of the transmission energy function are analyzed. Two offline and one online packet scheduling algorithms are proposed in Section \ref{sec: algorithm}. Simulation results and conclusions are given in Sections \ref{sec: simu} and  \ref{sec: conclusion}, respectively.

\section{System Model and Problem Formulation}\label{sec: sys model}

\subsection{System Model and Finite Blocklength Codes}\label{subsec: sys model}
\begin{figure}[!t]
  \centering
  \includegraphics[scale=0.9]{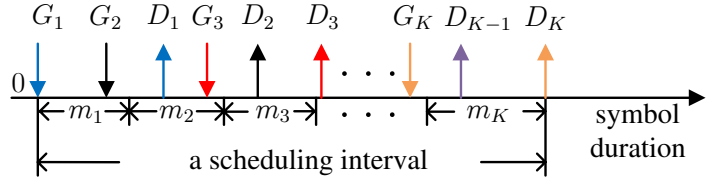}\\
  \caption{Delay constrained packet transmissions using finite blocklength codes with $m_{k}$ symbols for packet $k$}\label{General Case 1}
\end{figure}

We consider a packet scheduling problem where a transmitter schedules the transmission of
a sequence of $K$ data packets.
We assume long or medium range communications, where the energy
consumption is mainly contributed by data transmission and the circuit power consumption is negligible \cite{Miao-2009}.
The purpose is to minimize the expenditure of total transmission energy while satisfying the transmission deadline constraints imposed on the packets.
We assume that each packet $k$, which contains $N_k$ data bits, arrives at time $G_{k}$ and has to be completely delivered before time $D_{k}$ (see Fig. \ref{General Case 1}).
It is assumed that the scheduler obeys the FIFO rule \cite{Biyikoglu-2002, Chen-2007}, i.e., $G_k < G_{k+1}$ and $D_k < D_{k+1}$ for all $k$.
Without loss of generality, we set $G_{1} = 0$ and $G_{K+1} = D_{K}$, and assume $G_{k+1}<D_k$ for $k = 1,\ldots,K-1$ \cite{Chen-2007}.
The latter assumption implies that these $K$ packets belong to a ``scheduling interval", otherwise the scheduling problem can be decomposed into multiple independent scheduling problems. For example, if it happens that $G_{k+1} \geq D_{k}$ for some $k\in \{1,\ldots,K-1\}$, then the task of scheduling $K$ packets boils down to two independent problems, one for scheduling packet $1$ to packet $k$ and the other for scheduling the remaining packets.

To deliver the packets, the scheduler encodes the $N_k$ data bits of each packet $k$ into a codeword block (data payload of a packet) with a length of $m_k$ symbols, and transmits the codeword with a power $p_k > 0$. Note that rather than transmitting independent modulated symbols, each symbol in the codeword may depend on all $N_k$ data bits and correlate with other symbols in the same block.
We assume that some capacity-approaching coding strategy is used (e.g. low-density parity-check (LDPC) codes in \cite{Brink-2004}).
In the existing packet scheduling designs such as those in \cite{Biyikoglu-2002, Chen-2007, Chen-2009, Zafer-2009, Wang-2013}, it is assumed that the transmitted codeword can be successfully decoded by the receiver if $N_k$, $m_k$ and $p_k$ satisfy the Shannon channel capacity formula, i.e.,
\begin{align}\label{Shannon capacity}
\frac{N_{k}}{m_{k}} = \log_{2}(1+p_{k}h_{k}),
\end{align}
where $h_k \triangleq |\tilde{h}_{k}|^{2}$ and $\tilde{h}_{k}\in \mathbb{C}$ is a complex channel coefficient during transmitting packet $k$. In \eqref{Shannon capacity}, the noise power is assumed to be one.
It is important to note that, according to the Shannon's channel coding theorem \cite{Shannon-2001}, \eqref{Shannon capacity} is valid only when the block length $m_{k}$ approaches infinity \cite{Polyanskiy-2010, Gursoy-2013}. Obviously, under the deadline constraint, block length $m_k$ must be finite and satisfies $0 \leq m_k \leq D_k-G_k$. This implies that for delay-constrained applications, the existing designs \cite{Biyikoglu-2002, Chen-2007, Chen-2009, Zafer-2009, Wang-2013} that use \eqref{Shannon capacity} may fail to predict the true system performance.

Targeting at solving the aforementioned inaccuracy issue, the work \cite{Polyanskiy-2010} has generalized the Shannon capacity to the finite blocklength regime. It shows that, given a packet error probability $\epsilon_{k}\in (0,1)$, a transmission power $p_k$ and a block length $m_k$, 
the achievable data rate is given by
\begin{align}
   \frac{N_{k}}{m_{k}} &= \log_{2}(1+p_{k}h_{k})
    - \sqrt{\frac{1}{m_{k}}\left(1-\frac{1}{(p_{k}h_{k} + 1)^{2}}\right)}\frac{Q^{-1}(\epsilon_{k})}{\ln2}
    +\frac{\mathcal{O}(\log m_k)}{m_k}, \nonumber\\
    &\approx \log_{2}(1+p_{k}h_{k})
    - \sqrt{\frac{1}{m_{k}}\left(1-\frac{1}{(p_{k}h_{k} + 1)^{2}}\right)}\frac{Q^{-1}(\epsilon_{k})}{\ln2},
    \label{FBC rate}
\end{align}
where $Q(x) = \int_{x}^{\infty}\frac{1}{\sqrt{2\pi}}\exp(-t^{2}/2)dt$ is the Gaussian $Q$-function and $Q^{-1}$ denotes the inverse function of $Q$. The approximation in \eqref{FBC rate} is made by assuming that $m_k$ is no smaller than a threshold $\hat m>0$ so that the term $\frac{\mathcal{O}(\log m_k)}{m_k}$ becomes negligible.
Notably, according to \cite{Polyanskiy-2010}, \eqref{FBC rate} is valid even when the threshold $\hat m$ is as small as $100$. This is strongly contrast to the traditional Shannon capacity in \eqref{Shannon capacity} which is approximately true only if $\hat{m}\geq 10^5$ \cite{Brink-2004}. Therefore, \eqref{FBC rate} is particularly suitable for the packet scheduling problems with finite-blocklength codes.
However, this new formula, which is adopted in our work, is much more complicated than the Shannon capacity formula, as explained shortly.

It is worthwhile to notice that the new formula \eqref{FBC rate} reduces to \eqref{Shannon capacity} if one sets $\epsilon_{k} = 0.5$ in \eqref{FBC rate}, i.e., $Q^{-1}(0.5)=0$. In other words, any scheduling solutions $m_k$ and $p_k$ obtained based on \eqref{Shannon capacity} correspond to a packet error probability of $0.5$ in \eqref{FBC rate}. Therefore, the conventional designs in \cite{Biyikoglu-2002, Chen-2007, Chen-2009, Zafer-2009, Wang-2013} using \eqref{Shannon capacity} actually cannot guarantee reliable performance for finite-blocklength packet transmission.
In the next subsection, we use \eqref{FBC rate} to formulate an energy-efficient packet scheduling problem.

\subsection{Packet Scheduling Problem with Finite Blocklength Codes} \label{subsec: prob_formulation}

In this subsection, we formulate an offline packet scheduling problem by assuming that the packet arrival times $\{G_k\}$, deadlines $\{D_k\}$ and channel coefficients $\{h_k\}$ are known a priori.
The importance of studying the offline scheduling problem is twofold. First, offline solutions serve as performance lower bounds for an online algorithm. Second, offline solutions usually provide useful insights into the development of efficient online algorithms.
To formulate the problem, let us make the following assumption.

\begin{Assumption}\label{Ass1}
(No-idling assumption) The scheduler starts to transmit each packet $k$ right after the transmission of packet $(k-1)$ is complete, for $k=2,\ldots,K$.
\end{Assumption}
The no-idling assumption is intuitively justified as there is no benefit to delay the transmissions of packets, especially when the packets are subject to deadline constraints. Mathematically, the no-idling assumption is automatically satisfied if the transmission energy is a decreasing function of the blocklength $m_k$ and the scheduler targets at minimizing the transmission energy. As we will show shortly, this property is indeed true under some mild conditions on the packet blocklength $m_k$. Note that, under the no-idling assumption, the accumulated blocklength $\sum_{i=1}^k m_i$ represents the end time of the transmission of packet $k$ as well as the start time of the transmission of packet $(k+1)$.

We define the transmission energy of packet $k$ as $E_{k}(m_{k}, p_{k}) \triangleq m_{k}p_{k}$.
Under the FIFO rule and the no-idling assumption, we formulate the energy-efficient packet scheduling problem as follows
\begin{subequations}\label{P1}
\begin{align}
\!\!\!\!\!    \min_{\substack{p_k\geq0,m_k\geq 0,\\k=1,\ldots,K}} &~~  \sum_{k=1}^{K} E_{k}(m_{k}, p_{k}) \label{P1-target}\\
   \mbox{s.t.} &~~F_k(m_k,p_k)=0, ~\forall~k = 1,\ldots,K,\label{P1-0}\\
   &~~ m_{k} \geq \hat{m}, ~\forall~k = 1,\ldots,K, \label{P1-1}\\
   &~~ \textstyle\sum_{i=1}^{k} m_{i} \geq G_{k+1}, ~\forall~k = 1,\ldots,K, \label{P1-2}\\
   &~~ \textstyle\sum_{i=1}^{k} m_{i} \leq D_{k}, ~\forall~k = 1,\ldots,K, \label{P1-3}\\
   &~~ p_{k} \leq P_{\max}, ~\forall~k = 1,\ldots,K, \label{P1-4}
\end{align}
\end{subequations}
where
\begin{align}\label{eqn Fk}
 F_k(m_k,p_k)\triangleq \textstyle\sqrt{\frac{1}{m_{k}}\left(1-\frac{1}{(p_{k}h_{k} + 1)^{2}}\right)}\frac{Q^{-1}(\epsilon_{k})}{\ln2} - \log_{2}(1+p_{k}h_{k})+\frac{N_{k}}{m_{k}}
\end{align}
is a continuously differentiable function.

Problem \eqref{P1} aims to optimize the transmission power $p_k$ and blocklength $m_k$, in order to minimize the total transmission energy subject to the finite blocklength channel capacity formula \eqref{P1-0} and some scheduling constraints \eqref{P1-1}-\eqref{P1-4}.
Among the scheduling constraints, \eqref{P1-1} is the minimum blocklength constraint for \eqref{P1-0} holding true.
The constraints \eqref{P1-2} and \eqref{P1-3} are known as the \emph{causality constraint} and \emph{deadline constraint}, respectively \cite{Biyikoglu-2002,Chen-2007}. Specifically, for each packet $k$, \eqref{P1-2} indicates that the packet cannot be transmitted before its arrival time, while \eqref{P1-3} suggests that the transmission should be completely finished before its deadline.
Notice that \eqref{P1-2} and \eqref{P1-3} ensure\footnote{For each $k$, we have $G_{k} + m_{k} \leq \sum_{i=1}^{k-1} m_{i} + m_{k} = \sum_{i=1}^{k} m_{i} \leq D_{k}$. Thus, $m_k \leq D_k -G_k$.} $m_k \leq D_k -G_k$ for all $k$.
Equation \eqref{P1-4} indicates the maximum transmission power constraint.

Notably, by applying the implicit function theorem \cite{Krantz_Parks02} to $F_k(m_k,p_k)=0$ in \eqref{P1-0}, there exists a continuously differentiable function, denoted by $P_k$, such that $P_k(m_k) = p_k$.
Therefore, problem \eqref{P1} can be equivalently written as
\begin{subequations}\label{P2}
\begin{align}
   \min_{\substack{m_k\geq 0,\\ k=1,\ldots,K}} &~~\sum_{k=1}^{K}  E_{k}(m_{k},P_{k}(m_{k})) \\
   \mbox{s.t.}
   &~~ m_{k} \geq \hat{m}, ~\forall~k = 1,\ldots,K, \label{P2-1}\\
   &~~ \textstyle\sum_{i=1}^{k} m_{i} \geq G_{k+1}, ~\forall~k = 1,\ldots,K, \label{P2-2}\\
   &~~ \textstyle\sum_{i=1}^{k} m_{i} \leq D_{k}, ~\forall~k = 1,\ldots,K, \label{P2-3}\\
   &~~ P_{k}(m_{k}) \leq P_{\max}, ~\forall~k = 1,\ldots,K. \label{P2-4} 
\end{align}
\end{subequations}
We can see that $E_{k}(m_{k},P_{k}(m_{k}))$ is a function of $m_k$ only, thus in the sequel we use $E_{k}(m_{k})$ instead for brevity.

Let us examine problem \eqref{P2} by considering the traditional Shannon capacity formula in \eqref{Shannon capacity}.
In that case, the corresponding power function has the following closed-form expression
\begin{align}\label{P shannon}
P_{k}(m_{k})=\frac{1}{h_k}\left(2^{\frac{N_k}{m_k}}-1\right),
\end{align}
and the energy function is given by
\begin{align}\label{E shannon}
E_{k}(m_{k})= m_k P_{k}(m_{k}) = \frac{m_k\left(2^{\frac{N_k}{m_k}}-1\right)}{h_k}.
\end{align}
It can be shown that $P_{k}(m_{k})$ in \eqref{P shannon} is a monotonically decreasing function of $m_k$, and thus \eqref{P2-4} is equivalent to an explicit lower-bound constraint $m_k \geq \frac{N_{k}}{\log_{2}(1+P_{\max}h_{k})}$.
In addition, one can verify that $E_{k}(m_{k})$ in \eqref{E shannon} is a monotonically decreasing and convex function of $m_k$. Therefore, under the Shannon capacity formula, Assumption \ref{Ass1} is automatically satisfied. Moreover, problem \eqref{P2} is a convex optimization problem, which is efficiently solvable by off-the-shelf convex solvers (e.g., \texttt{CVX} \cite{cvx}).

Unfortunately, such monotonicity and convexity are no longer guaranteed when the finite-blocklength capacity formula \eqref{FBC rate} is considered. In fact, one even cannot obtain an explicit expression for the functions $P_{k}(m_{k})$ and $E_{k}(m_{k})$ under \eqref{FBC rate}. This implies that problem \eqref{P2} for the finite-blocklength case imposes a much greater challenge than the existing works in \cite{Chen-2007}.
To solve the problem, in the next section, we propose to characterize analytic conditions under which $E_{k}(m_{k})$ preserves desirable monotonicity and convexity. Later in Section IV, we further present two efficient optimization algorithms for handling problem \eqref{P2}.

\section{Monotonicity and Convexity of $E_{k}(m_{k})$ }\label{sec: mono and cvxit}
As one of the key results, the following proposition states the conditions for which $P_{k}(m_{k})$ and $E_{k}(m_{k})$ are monotonically decreasing.
\begin{Proposition}\label{L1}
Let $\tau_k \triangleq \frac{Q^{-1}(\epsilon_k)}{\sqrt{\hat{m}}}$ for packet error probability $\epsilon_k \in (0,0.5)$. It holds true that \\{\rm \bf (a)}
The power function $P_k(m_k)$ under \eqref{FBC rate} is decreasing for $m_k \geq \hat m$; \\
{\rm \bf (b)} The energy function $E_k(m_k)$ under \eqref{FBC rate} is decreasing for
\begin{align}\hat m\leq m_k
\leq g_{E_k} \triangleq \Xc_k^{-1}\left(-\frac{1}{\Wc\left(-\exp(-1-\frac{\tau_k}{2})\right)} - 1\right),
\end{align}
where $\Xc_k(m_k)$ is a function satisfying
\begin{align}
  \sqrt{\frac{1}{m_{k}}\left(1-\frac{1}{(\Xc(m_k) + 1)^{2}}\right)}\frac{Q^{-1}(\epsilon_{k})}{\ln2} - \log_{2}(1+\Xc(m_k))+\frac{N_{k}}{m_{k}}=0,
    \label{Xc}
\end{align}
and $\Wc(z)$ is the Lambert W function satisfying $\Wc(z)\exp(\Wc(z)) = z$ \cite{Corless-1996}.
\end{Proposition}

{\bf Proof:}
  Proposition \ref{L1} is proved by bounding the gradients of $P_k(m_k)$ and $E_k(m_k)$ which are obtained by applying the implicit function theorem \cite{Krantz_Parks02} to \eqref{P1-0}. The details are relegated to Appendix \ref{app1}.
\hfill $\blacksquare$

Proposition \ref{L1} has two significant aspects. First, according to the monotonic property of $P_k(m_k)$ in Proposition \ref{L1}(a), one can reformulate the implicit constraint \eqref{P2-4} as an explicit one.
Specifically, by Proposition \ref{L1}(a), \eqref{P2-4} is equivalent to bounding $m_k$ from the bottom, i.e., $m_k \geq \tilde m_k$, where $\tilde m_k=P_{k}^{-1}(P_{\max})$. 
From \eqref{P1-0},
one can show that $P_{k}^{-1}$ has a closed form as
\begin{align}\label{eq Pinv}
 P_{k}^{-1}(y) = &\bigg[
 \frac{1}{2\log_{2}(1+y h_{k})}\bigg(
\sqrt{1-\frac{1}{(y h_{k} +1)^{2}}}\frac{Q^{-1}(\epsilon_{k})}{\ln2}
  \notag \\
  &+ \sqrt{\bigg(1-\frac{1}{(y h_{k} +1)^{2}}\bigg)\bigg(\frac{Q^{-1}(\epsilon_{k})}{\ln2}\bigg)^2 - 4N_{k}\log_{2}(1+ y h_{k})}\bigg)
\bigg]^{2}.
\end{align}
Second, since from Proposition \ref{L1}(b) $E_k(m_k)$ is decreasing with $m_k$ for $m_k \leq g_{E_k}$, Assumption \ref{Ass1} is well justified if $m_k$ is restricted within that region.
Based on the discussions above, we rewrite problem \eqref{P2} as follows
\begin{subequations}\label{P3}
\begin{align}
\!\!\!\!\!    \min_{\substack{m_k\geq 0,\\ k=1,\ldots,K}} &~~\sum_{k=1}^{K}  E_{k}(m_{k}) \label{P3-0}\\
   \mbox{s.t.}&~~ \sum_{i=1}^{k} m_{i} \geq G_{k+1}, ~\forall k = 1,\ldots,K, \label{P3-1}\\
   &~~ \sum_{i=1}^{k} m_{i} \leq D_{k}, ~\forall k = 1,\ldots,K, \label{P3-2}\\
   &~~ m_{k} \geq \max\{\hat{m},\tilde{m}_{k}\},  ~\forall k = 1,\ldots,K,\label{P3-3}\\
      &~~ m_{k} \leq g_{E_k}, \forall k = 1,\ldots,K,  \label{P3-4}
\end{align}
\end{subequations}
where we have added constraints \eqref{P3-3} and \eqref{P3-4}.

We note that the constraints \eqref{P3-1} to \eqref{P3-4} are all linear. Therefore, what remains is to characterize the convexity of $E_{k}(m_{k})$.
This is established below.

\begin{Theorem}\label{T1}
Given $\tau_k \triangleq \frac{Q^{-1}(\epsilon_k)}{\sqrt{\hat{m}}} \in (0,\frac{\sqrt{3}}{3})$, the energy function $E_k(m_k)$ under \eqref{FBC rate} is a strictly convex function of blocklength $m_k$ for
\begin{align}
\hat m \leq m_k \leq g_{C,k} \triangleq
\textstyle \Xc_k^{-1}\left(\exp\left(\eta(\tau_k) + \frac{\tau_k}{2}\right) -1\right),
\end{align}
where
\begin{align}
 \eta(\tau_k) \triangleq \frac{3 + \sqrt{9 + 12\tau_k(1-\sqrt{3}\tau_k)}}{4(1-\sqrt{3}\tau_k)}.
\end{align}
\end{Theorem}
{\bf Proof:}
See Appendix \ref{app3}.
\hfill $\blacksquare$

Theorem \ref{T1} implies that problem \eqref{P3} is a convex optimization problem if $m_k \leq g_{C,k}$.
It will be seen in the next section that problem \eqref{P3} can be globally solved by a low complexity algorithm as long as it is a (strictly) convex problem. In the general case for which problem \eqref{P3} may not be convex, we also present an efficient approximation algorithm.
Before ending this section, we have two remarks regarding $g_{E_k}$, $g_{C,k}$ and $\tau_k$.

\begin{figure}[!t]
  \centering
  \subfigure[]{
    \label{upper bound epsilon} 
    \includegraphics[width=0.48\linewidth]{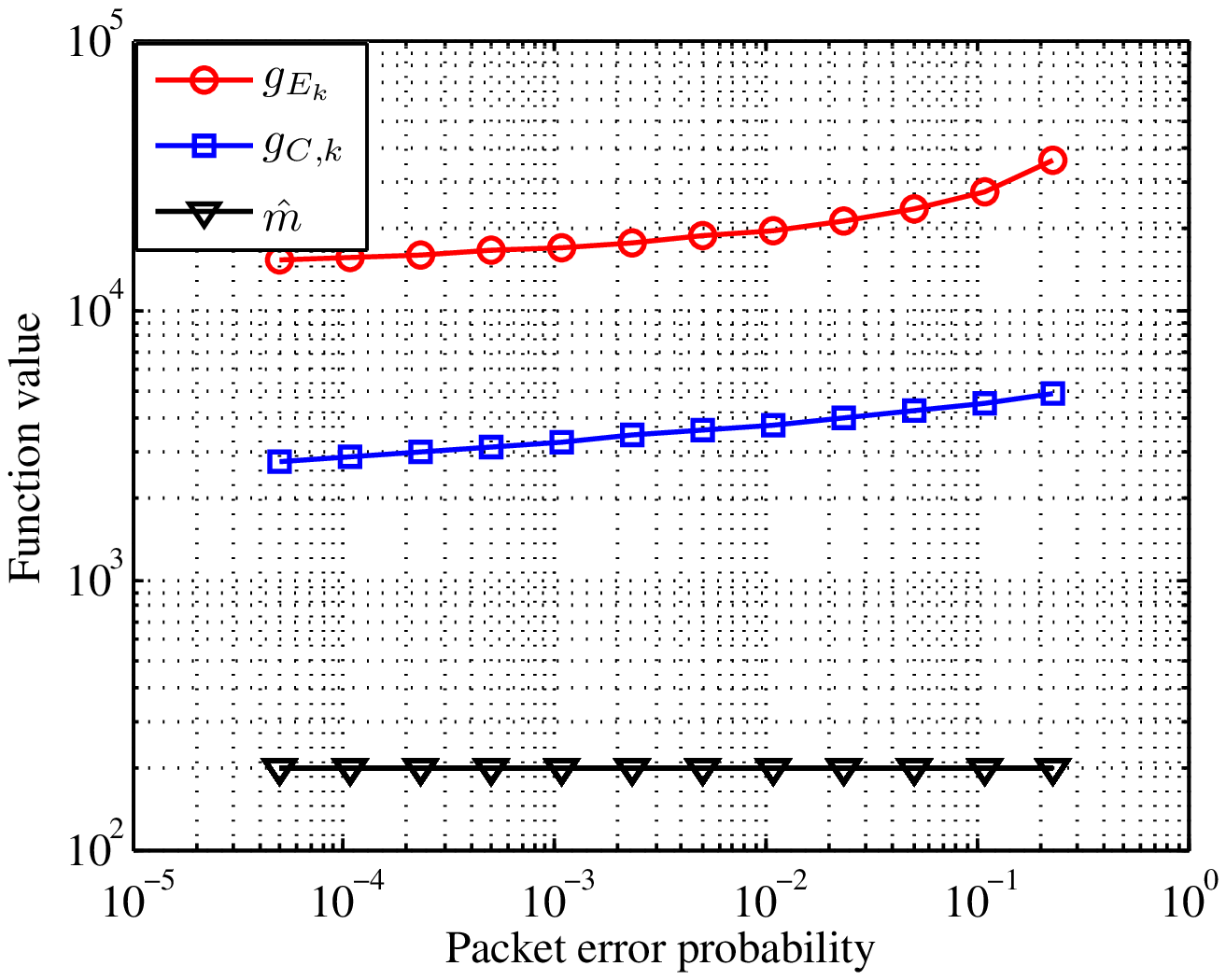}}
  \subfigure[]{
    \label{upper bound N} 
    \includegraphics[width=0.484\linewidth]{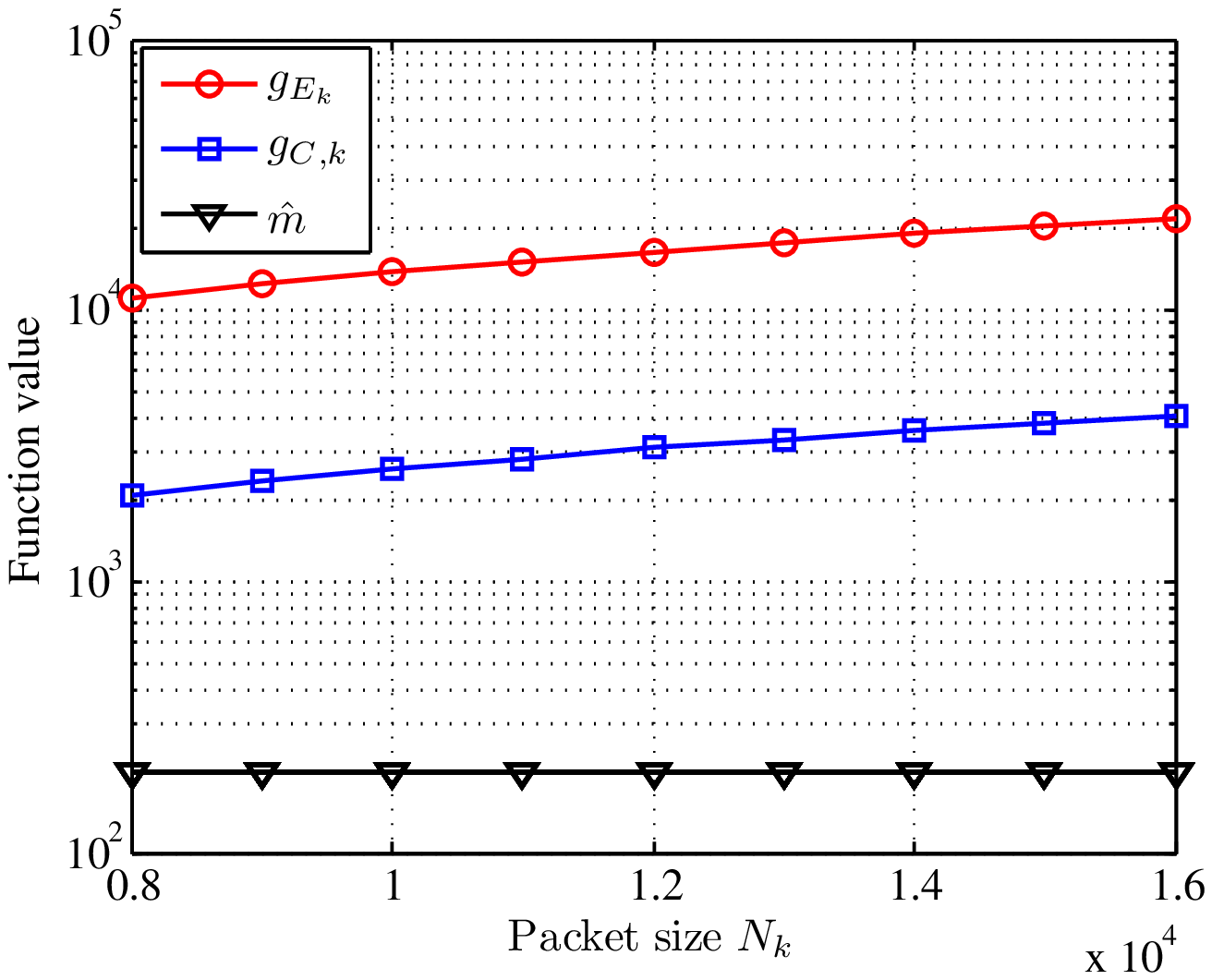}}
  \caption{Upper bounds $g_{E_k}$ and $g_{C,k}$ versus $\epsilon_k$ and $N_k$;
  (a) $N_k=1.2\times 10^4$, $h_k=10$, and $\hat{m}=200$; (b) $\epsilon= 5\times 10^{-4}$, $h_k=10$, and $\hat{m}=200$.}
  \label{upper bound}
\end{figure}

\begin{Remark}\label{R1}{\rm
It is worthwhile to note that
the upper bounds $g_{E_k}$ and $g_{C,k}$ characterized by the function $\Xc_k$ in \eqref{Xc} are independent of the channel gain $h_k$. Moreover, the upper bounds $g_{E_k}$ and $g_{C,k}$ are arguably large enough, and in general do not have an impact on the solutions to problem \eqref{P3}.
To see this, we numerically draw the curves of $g_{E_k}$ and $g_{C,k}$ with respect to $\epsilon_k$ in Fig. \ref{upper bound epsilon} and with respect to $N_k$ in Fig. \ref{upper bound N}, respectively.
One can see from the two figures that the gaps between
$\hat m$ and $g_{E_k}$ and $g_{C,k}$ are fairly large (e.g., for $N_k=1.2\times 10^4$ and $\epsilon_k = 5\times 10^{-4}$, we have $g_{E_k}=1.5\times10^4$ and $g_{C,k}=3\times 10^3$). This implies that, for the scenarios
with small values of $D_k-G_k \leq \min \left\{g_{C,k},g_{E_k}\right\}$ for all $k$,
the constraint \eqref{P3-4} is automatically satisfied and problem \eqref{P3} is a convex optimization problem.
}
\end{Remark}

\begin{Remark}\label{R2}{\rm
The condition of $\tau_k \in (0,\frac{\sqrt{3}}{3})$ in Theorem \ref{T1} is minor. Specifically, one can readily show that $\tau_k=\frac{Q^{-1}(\epsilon_k)}{\sqrt{\hat{m}}} \in (0,\frac{\sqrt{3}}{3})$ is equivalent to $\epsilon_k \in (Q(\frac{\sqrt{3\hat{m}}}{3}), 0.5)$.
For $\hat m = 200$, the condition $\tau_k \in (0,\frac{\sqrt{3}}{3})$ can be satisfied whenever $\epsilon_k \in (1.6 \times 10^{-16}, 0.5)$, which is true in practical scenarios.
}
\end{Remark}

\section{Offline and Online Scheduling Algorithms}\label{sec: algorithm}
In the previous section, we have shown that the transmission energy function using finite blocklength codes in \eqref{P3-0} can still preserve the monotonicity and convexity under mild conditions on the code blocklength.
In this section, we first present a feasibility condition of problem \eqref{P3}. Then, we respectively propose two efficient offline algorithms for solving problem \eqref{P3} in Section \ref{subsec: MLWF} and \ref{subsec: SUM}. Finally, an online algorithm is proposed in Section \ref{subsec: online}.

\subsection{Feasibility Condition}\label{subsec: feasibility}
Due to the deadline constraint, problem \eqref{P3} may not have a feasible solution.
In that case, the mechanism of admission control may be adopted, which however is beyond the scope of the current paper.
Here we present a sufficient condition for which problem \eqref{P3} is feasible.
\begin{Proposition}\label{L2}
Suppose that the packet arrival times $\{G_k\}$ and deadlines $\{D_k\}$ satisfy
\begin{subequations}\label{sufficient condition}
\begin{align}
   &~~ G_{k+1} - G_{k} \geq \max\{\hat{m},\tilde{m}_{k}\}, ~\forall k = 1,\ldots,K, \label{sufficient condition-1}\\
   &~~ D_{k} - G_{k} \leq g_{E_k}, ~\forall k = 1,\ldots,K. \label{sufficient condition-2}
\end{align}
\end{subequations}
Then problem \eqref{P3} has a feasible solution.
\end{Proposition}
{\bf Proof:}
We show that $\bar m_{k} = G_{k+1} - G_{k}$, $k=1,\ldots,K$, is a feasible solution to problem \eqref{P3} given \eqref{sufficient condition}. First, since $G_{k+1} < D_k$ for all $k$ from the ``same scheduling interval" assumption in Section \ref{subsec: sys model}, \eqref{sufficient condition-1} and \eqref{sufficient condition-2} respectively imply that $\{\bar m_{k}\}$ satisfies \eqref{P3-3} and \eqref{P3-4}.
Second, note that $\sum_{i=1}^{k} \bar m_i = G_{k+1} < D_{k}$. So, $\{\bar m_{k}\}$ satisfies the constraint \eqref{P3-1} with equality and also satisfies the constraint \eqref{P3-2}.
\hfill $\blacksquare$

By assuming that problem \eqref{P3} is feasible (e.g., the feasibility condition \eqref{sufficient condition} holds), we next study efficient algorithms for solving problem \eqref{P3}. We remark that solving \eqref{P3} is challenging due to the following two reasons.
First, problem \eqref{P3} is in general not convex without adding the additional constraints $m_k\leq g_{C,k},\forall k$ from Theorem \ref{T1}. Second, even under conditions for which problem \eqref{P3} is convex, standard optimization tools (such as \texttt{CVX} \cite{cvx}) cannot be employed to solve problem \eqref{P3} since the energy function $E_k(m_k)$ does not have an explicit, close-form expression. In the next two subsections, we respectively present two efficient offline algorithms for overcoming these difficulties.

\subsection{Multi-level Watering-Filling Algorithm for Convex Packet Scheduling}\label{subsec: MLWF}
In this subsection, we assume that problem \eqref{P3} is a (strictly) convex optimization problem.
For example, this can be guaranteed if the constraint $m_k\leq g_{C,k}, \forall k$ is imposed, i.e., \eqref{P3-4} is replaced with $m_{k} \leq \min\{g_{E_k},g_{C,k}\}, \forall k$. As aforementioned, standard convex solvers cannot be used for solving problem \eqref{P3}. Fortunately, the multi-level water-filling (MLWF) algorithm proposed in \cite{Wang-2013} can be modified to solving problem \eqref{P3} without the need of explicit expressions of $E_k(m_k)$. Note that in \cite{Wang-2013} the MLWF algorithm was proposed to solve a rate-controlled packet scheduling problem assuming a strictly convex energy function. While \cite{Wang-2013} has considered a different system setup, the optimization problem studied therein happens to have a similar form as problem \eqref{P3} and thereby the MLWF algorithm can be modified to solve problem \eqref{P3}.

To briefly illustrate how to apply the MLWF algorithm for problem \eqref{P3}, let us denote $\{m_{k}^\ast\geq 0,k = 1,\ldots,K\}$ as the optimal solution to problem \eqref{P3}, and denote $\{\mu_{k}^\ast\geq 0, k = 1,\ldots,K\}$ and $\{\lambda_{k}^\ast\geq 0, k = 1,\ldots,K\}$ as the optimal Lagrange multipliers associated with the constraint \eqref{P3-1} and \eqref{P3-2}, respectively. Then, $\{m_{k}^\ast,\mu_{k}^\ast,\lambda_{k}^\ast\}$ satisfy the following Karush-Kuhn-Tucker (KKT) optimality conditions \cite{Boyd-2009} of problem \eqref{P3}:
\begin{align}
  m_{k}^{\ast} = \min\{\ell_k,\max\{\phi_{k}(\omega_{k}),u_k\}\},&~\forall k = 1,\ldots,K,
  \label{KKT}\\
  \mu_{k}^{\ast}\left(\sum_{i=1}^{k} m_{i}^{\ast} - G_{k+1}\right)=0, &~\forall k = 1,\ldots,K, \label{slackness-1}\\
  \lambda_{k}^{\ast}\left(\sum_{i=1}^{k} m_{i}^{\ast} - D_{k}\right)=0, &~\forall k = 1,\ldots,K, \label{slackness-2}
\end{align}
where $\ell_k\triangleq \max\{0,\hat{m},\tilde{m}_{k}\}$, $u_k\triangleq \min\{g_{E_k},g_{C,k}\}$,
\begin{equation}\label{omega}
\omega_{k} \triangleq \sum_{i=k}^{K} \left(\mu_{i}^{\ast} - \lambda_{i}^{\ast}\right),
\end{equation}
and $\phi_{k}$ is the the inverse function of $E_{k}'$ (the gradient of $E_k$). Clearly, once the ``water levels" $\omega_k,$ $k=1,\ldots,K$, are obtained, the optimal packet length $\{m_k^\ast\}$ can be evaluated through \eqref{KKT}. It is shown in \cite{Wang-2013} that, by carefully exploiting the structure of problem \eqref{P3}, $\omega_k,$ $k=1,\ldots,K$, can be determined by a low-complexity search algorithm. Due to the space limit, we refer the readers to \cite[Section III]{Wang-2013} for the details.

However, we emphasize here that, to implement the MLWF algorithm, one must be able to evaluate the function value of $\phi_k(\omega_k)$ in \eqref{KKT}.
This task is non-trivial for our problem \eqref{P3}.
When the Shannon capacity formula in \eqref{Shannon capacity} is used, it can be shown that $\phi_k(\omega_k)$ has a closed form
$\phi_{k}(\omega_{k}) = \frac{N_k \ln2}{1+\Wc\left(-\frac{\omega_k h_k +1}{e}\right)}$,
where $\Wc$ is the Lambert W function and $e$ is the Euler's number. However, when the finite blocklength channel capacity in \eqref{FBC rate} is used, $\phi_k(\omega_k)$ no longer has a close-form expression. Fortunately, based on the monotonicity of $P_k(m_k)$ and $E_k'(m_k)$ as proved in Proposition \ref{L1} and Theorem \ref{T1}, we are able to evaluate $\phi_k(\omega_k)$ via bisection search.
Note that $\phi_{k}$ is monotonically increasing as $E_{k}'$ is monotonically increasing when $E_k$ is strictly convex.
Specifically, we present the bisection algorithm for evaluating $\phi_{k}(\omega_k)$ in Algorithm \ref{a-01}. In Step 4, we can evaluate $E_k'(\bar m_k)$ for a given $\bar m_k$ by the following formula
\begin{align}
  E_{k}'(m_{k}) &= P_{k}(m_{k}) + m_{k}P_{k}'(m_{k}) \nonumber\\
   & = P_{k}(m_{k}) - m_{k}\bigg(\frac{\partial F_{k}(m_{k},p_{k})}{\partial m_{k}}\bigg |_{p_k=P_k(m_k)}\bigg)
\left(\frac{\partial F_{k}(m_{k},p_{k})}{\partial p_{k}} \bigg |_{p_k=P_k(m_k)}\right)^{-1},\label{E der}
\end{align}
where the second equality is obtained by applying the implicit function theorem \cite{Krantz_Parks02} to $F_{k}(m_{k},p_{k})$ $=0$ (see \eqref{P1-0}).
Explicit expressions of $\frac{\partial F_{k}(m_{k},p_{k})}{\partial m_{k}}$ and $\frac{\partial F_{k}(m_{k},p_{k})}{\partial p_{k}}$
are given in \eqref{app1-5-1} and \eqref{app1-6-1} in Appendix \ref{app1}, respectively.
As seen, it remains to calculate $P_k(m_k)$. This again can be achieved by bisection search based on the monotonicity of $P_k(m_k)$, as we show in Algorithm \ref{a-02}.

\begin{algorithm}[!thb]
\caption{Bisection algorithm for evaluating $\phi_{k}(\omega_k)$}\label{a-01}
\begin{algorithmic}[1]
\STATE {\bf Given} initial values of $m_{u}=D_k-G_k$ and $m_{\ell}=0$, and the accuracy $\varepsilon_1$.
\WHILE{$m_{u}-m_{\ell} > \varepsilon_1$}
\STATE{$\bar m_k\leftarrow  {1 \over 2} (m_{u}+m_{\ell})$}
\STATE{Calculate $E_{k}'(\bar m_k)$ by \eqref{E der}, where $P_k(m_{k})$ can be obtained by Algorithm \ref{a-02}}
\IF{$E_{k}'(\bar m_k) > \omega_k$}
\STATE{$m_{u}\leftarrow \bar m_k$}
\ELSE
\STATE{$m_{\ell}\leftarrow \bar m_k$}
\ENDIF
\ENDWHILE
\RETURN{$\phi_{k}(\omega_k)\leftarrow \bar m_k$}
\end{algorithmic}
\end{algorithm}

\begin{algorithm}[!thb]
\caption{Bisection algorithm for evaluating power function $P_{k}(m_k)$}\label{a-02}
\begin{algorithmic}[1]
\STATE {\bf Given} initial values of $P_u=P_{\max}$ and $P_\ell=0$, and the accuracy $\varepsilon_2$.
\WHILE{$P_u-P_\ell > \varepsilon_2$}
\STATE{$\bar p_k\leftarrow {1 \over 2} (P_u+P_\ell)$}
\STATE{Calculate $\bar m_k$ $\leftarrow P_k^{-1}(\bar p_k)$ by \eqref{eq Pinv}.}
\IF{$\bar{m}_k < m_k$}
\STATE{$P_u\leftarrow \bar p_k$}
\ELSE
\STATE{$P_\ell\leftarrow \bar p_k$}
\ENDIF
\ENDWHILE
\RETURN{$P_{k}(m_k)\leftarrow \bar p_k$}
\end{algorithmic}
\end{algorithm}

The computational complexity of the MLWF algorithm is given as follows. By \cite{Wang-2013}, given a total number of $K$ packets,
the MLWF algorithm requires  $\mathcal{O}(K)$ rounds of search to attain the global optimal solution.
In each round of the MLWF algorithm, one needs to evaluate the function values of $\phi_k(\omega_k), k=1,\ldots,K$, by  Algorithm 1, and in each iteration of Algorithm  1 one has to run Algorithm 2 for obtaining  $P_k(m_k)$.
Algorithm  1 is a bisection search method which is known to have a complexity order $\log_{2}(\frac{D_k - G_k}{\varepsilon_1})$; similarly, the complexity order of Algorithm 2 is $\log_{2}(\frac{P_{\rm max}}{\varepsilon_2})$.
Thus, the MLWF algorithm has a total complexity of $\mathcal{O}(K^2\log_{2}(\frac{\max_k (D_k - G_k)}{\varepsilon_1})\log_{2}(\frac{P_{\rm max}}{\varepsilon_2}))$.

\subsection{Non-Convex Packet Scheduling Based on Successive Upper-bound Minimization}\label{subsec: SUM}
In the absence of constraints $m_k\leq g_{C,k}$, $k=1,\ldots,K$, problem \eqref{P3}  in general is not a convex problem. In that case, we propose to solve problem \eqref{P3} by the successive upper-bound minimization (SUM) method \cite{Razaviyayn-2013BSUM}.
Specifically, the SUM method for solving problem \eqref{P3} involves solving a sequence of the following subproblems: for iterations $r=0,1,2,\ldots,$
\begin{subequations}\label{PSUM}
\begin{align}
\!\!\!\!\!  \{m_k^{r+1}\}_{k=1}^K=\arg~  \min_{\substack{m_k\geq 0,\\ k=1,\ldots,K}} &~~\sum_{k=1}^{K}  U_{k}(m_{k}; m_k^r) \\
   \mbox{s.t.}&~~ \text{constraints~} \eqref{P3-1} - \eqref{P3-4},
\end{align}
\end{subequations}
where
\begin{align}\label{ub function}
  U_{k}(m_{k}; m_k^r) \triangleq E_{k}(m_{k}^{r})
  + E_{k}'(m_{k}^{r})(m_{k}-m_{k}^{r}) + \frac{\gamma}{2}|m_{k}-m_{k}^{r}|^{2}
\end{align}
is a proximal first-order approximation of $E_k(m_k)$ around $m_{k}^{r}$ and $\gamma>0$ is a penalty parameter; see Algorithm \ref{a-SUM}.

The SUM method has several advantages. Firstly, the quadratic objective function $U_{k}(m_{k}; m_k^r)$ can be explicitly computed since $E_{k}(m_{k}^{r}) = m_{k}^{r}P_{k}(m_{k}^{r})$ and $E_{k}'(m_{k}^{r})$ can be respectively evaluated by Algorithm \ref{a-02} and \eqref{E der}.
This method therefore avoids handling problem \eqref{P3} directly where $E_k(m_k)$ has no closed-form expression.
Secondly, the objective function $U_{k}(m_{k}; m_k^r)$ is strictly convex with respect to $m_k$, and therefore the standard convex solvers or even the MLWF algorithm in the previous subsection can be used to solve problem \eqref{PSUM} efficiently.
Finally, in contrast to the MLWF algorithm which relies on the specific problem structure of \eqref{P3}, the SUM method is more flexible in the sense that it can be extended to handle more complex scheduling constraints.

According to \cite{Razaviyayn-2013BSUM}, when $E_k'(m_k)$ is Lipschitz continuous and for a large enough $\gamma$, Algorithm \ref{a-SUM} is guaranteed to converge to the set of stationary points of problem \eqref{P3}. Moreover, if problem \eqref{P3} is convex, Algorithm \ref{a-SUM} converges to the global optimal solution set.
According to \cite{Hong-2013}, the iteration number of Algorithm \ref{a-SUM} is $\mathcal{O}(\frac{1}{\varepsilon})$, where $\varepsilon$ is the desired solution accuracy.
When the MLWF algorithm is used to solve the subproblem \eqref{PSUM} (which requires a complexity of  $\mathcal{O}( K^2 \log_{2}(\frac{P_{\rm max}}{\varepsilon_2}))$), Algorithm \ref{a-SUM} has a complexity of $\mathcal{O}(\frac{K^2}{\varepsilon}\log_{2}(\frac{P_{\rm max}}{\varepsilon_2}))$.

\begin{algorithm}[!thb]
\caption{SUM method for solving problem \eqref{P3}}\label{a-SUM}
\begin{algorithmic}[1]
\STATE Set $r=0$. Given a set of feasible $\{m_{k}^{0}\}_{k=1}^K$ and desired accuracy $\varepsilon$.
\REPEAT
\STATE{Calculate $U_{k}(m_{k}; m_k^r)$ in \eqref{ub function} by Algorithm \ref{a-02} and \eqref{E der}.}
\STATE{Solve \eqref{PSUM} by standard convex solvers or the MLWF algorithm, and obtain $\{m_k^{r+1}\}_{k=1}^K$.}
\STATE{Set $r \leftarrow r+1$.}
\UNTIL{$\sum_{k=1}^K |m_k^{r}-m_k^{r-1}|^2 < \varepsilon$.}
\end{algorithmic}
\end{algorithm}

\subsection{Rolling-Window Based Online Scheduling Algorithm}\label{subsec: online}

In the previous two subsections, offline scheduling algorithms are developed by assuming that the scheduler has full knowledge of the arrival times and deadlines of all the packets. However, in practical situations, only the information of arrived packets are known and an online algorithm that can perform real-time scheduling is desired. In this subsection, we present a rolling-window based online scheduling algorithm for such a purpose.

\begin{figure}[!h]
\begin{center}
  \centering
  \includegraphics[scale=0.85]{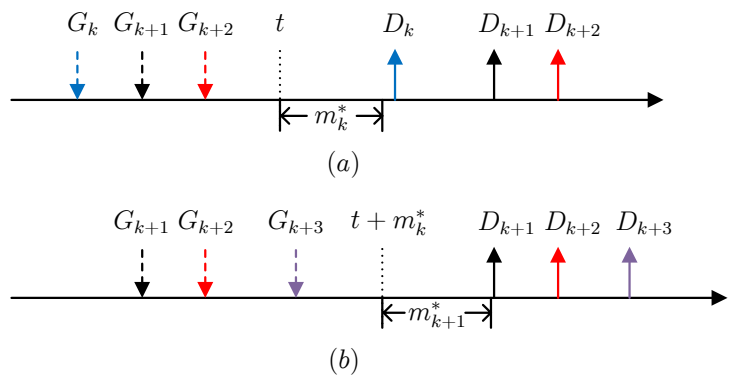}\\
  \caption{Example for illustrating the proposed online scheduling algorithm.}\label{OnlineCase}
  \end{center}
\end{figure}

The idea of the proposed online algorithm is to always schedule the most urgent packet at the current time for transmission, while taking into account the scheduling constraints of the packets that have arrived at and before the current time. To illustrate this online strategy, let us consider the example in Fig. \ref{OnlineCase}.
As shown in Fig. \ref{OnlineCase}(a), suppose that, at current time $t$, the packet $k$, packet $k+1$ and packet $k+2$ have arrived but not been scheduled yet.
Then, the scheduler considers a scheduling window from the current time to $D_{k+2}$. Note that, given the arrived packets, this is the longest scheduling window one can choose at current time. An obvious advantage of this strategy is that the deadline information of all arrived packets can be taken into account in the optimization which avoids myopic scheduling solutions.

Specifically, the scheduler sets the arrival times of the three packets to zero, and sets their deadlines to $D_{k} - t$, $D_{k+1} - t$ and $D_{k+2} - t$, respectively. Then, the scheduler applies the offline scheduling algorithm (e.g., the proposed algorithms in Section \ref{subsec: MLWF} and \ref{subsec: SUM}) to the scheduling window\footnote{Similar to \cite{Wang-2015}, in the online case, the future channel coefficients are assumed known or can be estimated through channel prediction methods such as Kalman filtering \cite{Shikur-2015}.} and obtains the optimal packet lengths $m_{k}^{\ast}, m_{k+1}^{\ast}$ and $m_{k+2}^{\ast}$, respectively. The scheduler, however, discards $m_{k+1}^{\ast}$ and $m_{k+2}^{\ast}$ and applies $m_{k}^{\ast}$ to the transmission of packet $k$ only.
At time $t+m_{k}^{\ast}$, when the transmission of packet $k$ finishes, the scheduler then repeats the optimization by moving the window forward which contains all the arrived and unscheduled packets (e.g., packets $k+1,$ $k+2$ and $k+3$ shown in Fig. \ref{OnlineCase}(b)).
Note that the complexity of the online algorithm depends on the offline algorithm used and the number of packets appearing in each scheduling window.
We will show in the next section that the proposed online scheduling algorithm outperforms some myopic strategies.

\section{Simulation Results}\label{sec: simu}

\subsection{Simulation Setup}
In the simulations, we assume that the $K$ packets contain equal size
of $N_{k} = 1.2 \times 10^{4}$ bits. The block fading channel
follows Rayleigh distribution with a scale parameter $\sigma$.
The packet error probabilities for all packets are set the same, i.e.,
$\epsilon\triangleq \epsilon_1=\cdots=\epsilon_K$. The maximum transmission power $P_{\max}$  is set to $26$ dBW.
Note that in the considered energy-efficient problem, the maximum power $P_{\max}$
may not always be used up, i.e., $p_k<P_{\max}.$
The transmission energy of each packet $k$ is $m_k T_s p_k$, where the symbol
duration $T_s$ is set to $66.7~\mu$s following the LTE standard
\cite{Sesia-2009}.

The packet arrival interval $(G_{k+1} - G_{k})$ is
generated according to a truncated exponential distribution with
mean $\nu\hat{m}$ in the time window $[\nu -1, \nu
+1]\times\hat{m}$ \cite{Jin-2012}. Similarly, the lifetime
of packet $(D_{k} - G_{k})$ follows a truncated exponential
distribution with mean $n\hat{m}$ in the time window $[n -1,n
+1]\times\hat{m}$. In general, a smaller value of $\nu$
indicates a smaller packet arrival interval while a smaller
value of $n$ implies a more stringent delay constraint for
scheduling. One can check that if $3 < \nu \leq n-2$, then the
FIFO rule (i.e., $D_{k}<D_{k+1}~\forall k$) and the scheduling interval assumption (i.e., $G_{k+1}<D_k~\forall k$) are both satisfied. Moreover, if $n +1
\leq \left\lfloor \min_{k=1,\ldots,K}
\left\{g_{C,k},g_{E_k}\right\}/\hat{m}\right\rfloor$, then $m_k\leq \min\{g_{C,k},g_{E_k}\}~\forall k$, that is,
problem \eqref{P3} must be a strictly convex problem.
All the simulation results to be presented are obtained by averaging over $100$ independent channel realizations and $100$ independent packet generations per channel realization.

\subsection{Energy Underestimation due to Shannon Capacity Formula}

\begin{figure}[t]
  \centering
  \subfigure[]{
    \label{EC-Epsilon-nu6} 
    \includegraphics[width=0.48\linewidth]{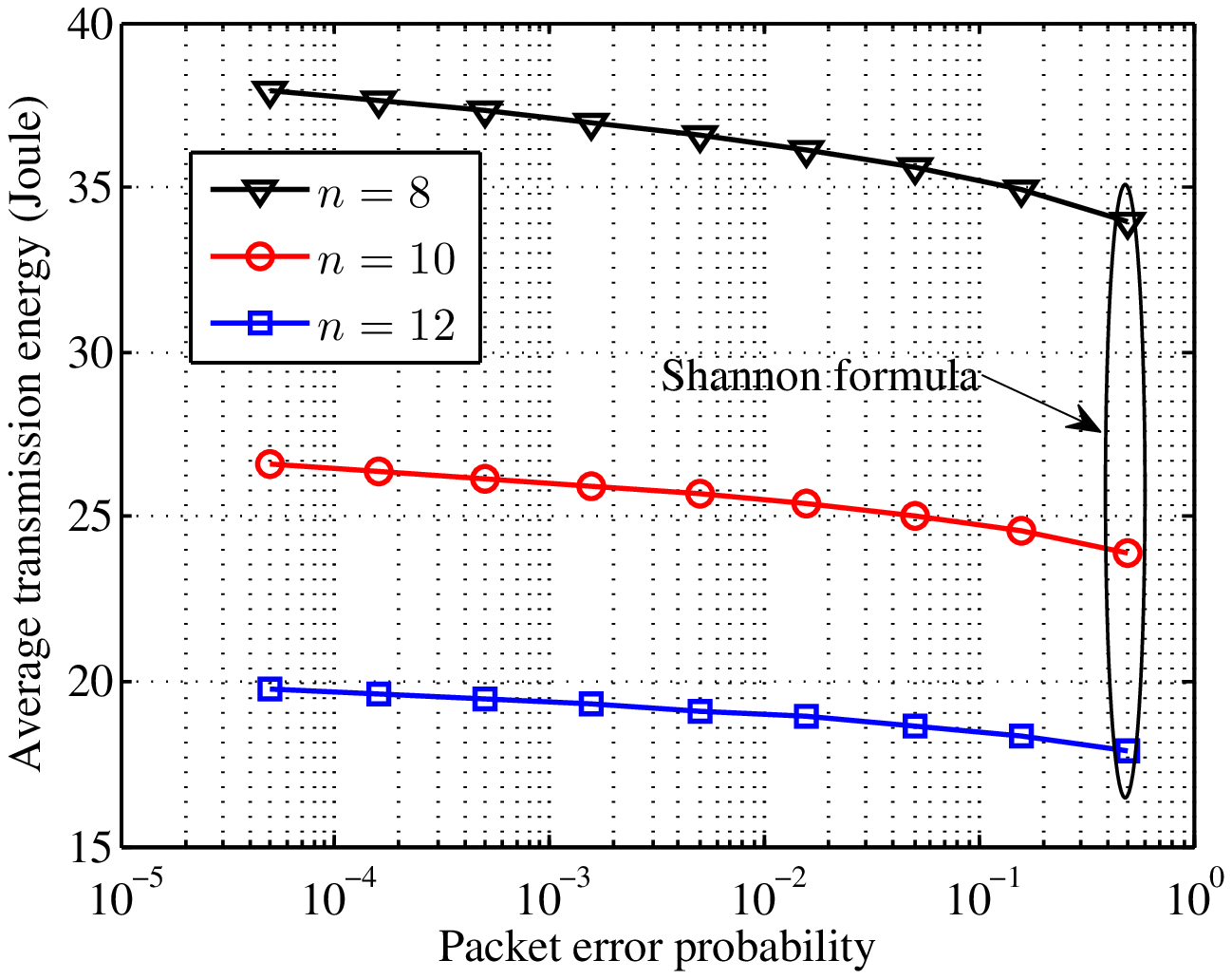}}
  \subfigure[]{
    \label{EU-Epsilon-nu6} 
    \includegraphics[width=0.48\linewidth]{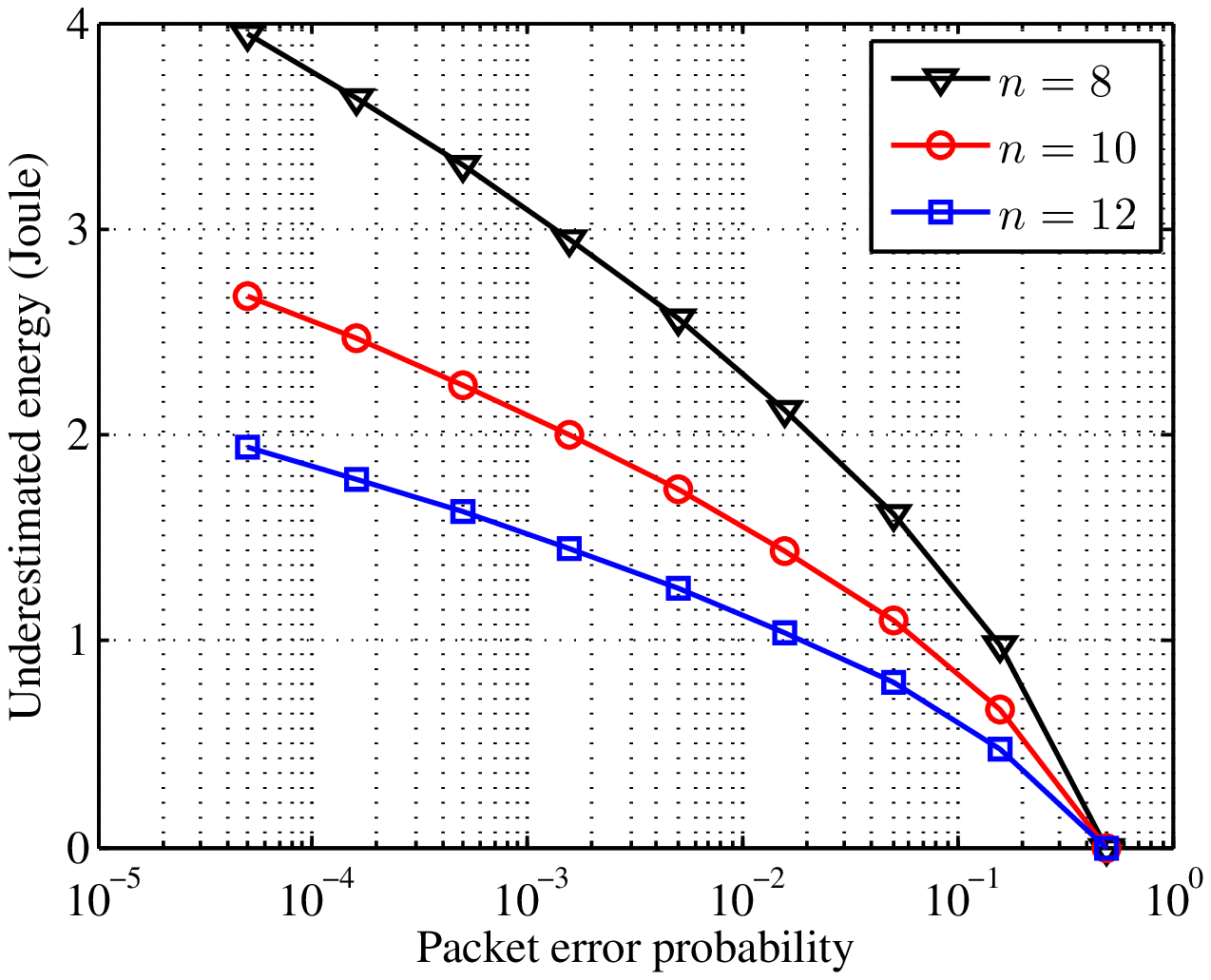}}
  \caption{Transmission energy and underestimated energy vs. packet error probability, for $\nu = 6$, $K =5$, $\sigma = 10$ and different values of $n$ (packet lifetime).}
\end{figure}

Let us first examine the scheduling performance of problem \eqref{P3}.
{The MLWF algorithm discussed in Section \ref{subsec: MLWF} is used to solve problem \eqref{P3}}.
Fig. \ref{EC-Epsilon-nu6} displays the average transmission energy
versus packet error probability $\epsilon$, under different
values of packet lifetime parameters $n$. The controlling
parameter $\nu$ for packet arrival interval is set to six.
As discussed in Section \ref{subsec: sys model},
the traditional design based on the Shannon capacity
formula corresponds to problem \eqref{P3} with
$0.5$ packet error probability. From curves in Fig.
\ref{EC-Epsilon-nu6}, we can see that the traditional design based on the Shannon capacity
formula suffers significant
energy underestimation. For example, from the curve with $n=10$,
the average transmission energy of the considered problem \eqref{P3} is $26.17$ Joule for $\epsilon=5\times10^{-4}$,
whereas the transmission energy obtained by the traditional design
($\epsilon=0.5$) is $23.93$ Joule. Thus, the traditional design
underestimates a total $2.24$-Joule transmission energy (about
10$\%$) for achieving $5\times10^{-4}$ packet error probability.
Indeed, from Fig. \ref{EC-Epsilon-nu6}, energy underestimation
always exists for $\epsilon < 0.5$ and the amount of
underestimated energy increases when the packet error probability decreases.
We can also observe from Fig. \ref{EC-Epsilon-nu6} that a
larger value of $n$ results in less transmission energy
consumption. This is because larger $n$ implies less stringent
delay constraint and a longer code blocklength is allowed according to our simulation setup.
By the fact that the energy function is decreasing with the blocklength (see Proposition \ref{L1}),
less energy is consumed for larger value of $n$.

To emphasize the energy underestimation issue due to the Shannon capacity formula, we further show in Fig.
\ref{EU-Epsilon-nu6} the amount of underestimated energy in Fig.
\ref{EC-Epsilon-nu6} versus the packet error probability. One can observe that the energy
underestimated by the traditional design using Shannon capacity
formula could be significant, especially when the desired packet
error probability is small and the delay constraint is stringent
(small $n$). Specifically, for $n=8$, $n=10$ and $n=12$, the
amount of underestimated energy can respectively be up to
$10.4\%$, $10.1\%$ and $9.8\%$ of the predicted energy from
problem \eqref{P3} using new capacity formula.
These simulation results not only confirm the intuition that energy consumption is increasing with lower latency and higher communication reliability, but also well demonstrate the necessity of the
finite-blocklength channel capacity for energy-efficient packet scheduling.

\begin{figure}[t]
  \centering
  \subfigure[]{
    \label{EC-Epsilon-n10} 
    \includegraphics[width=0.48\linewidth]{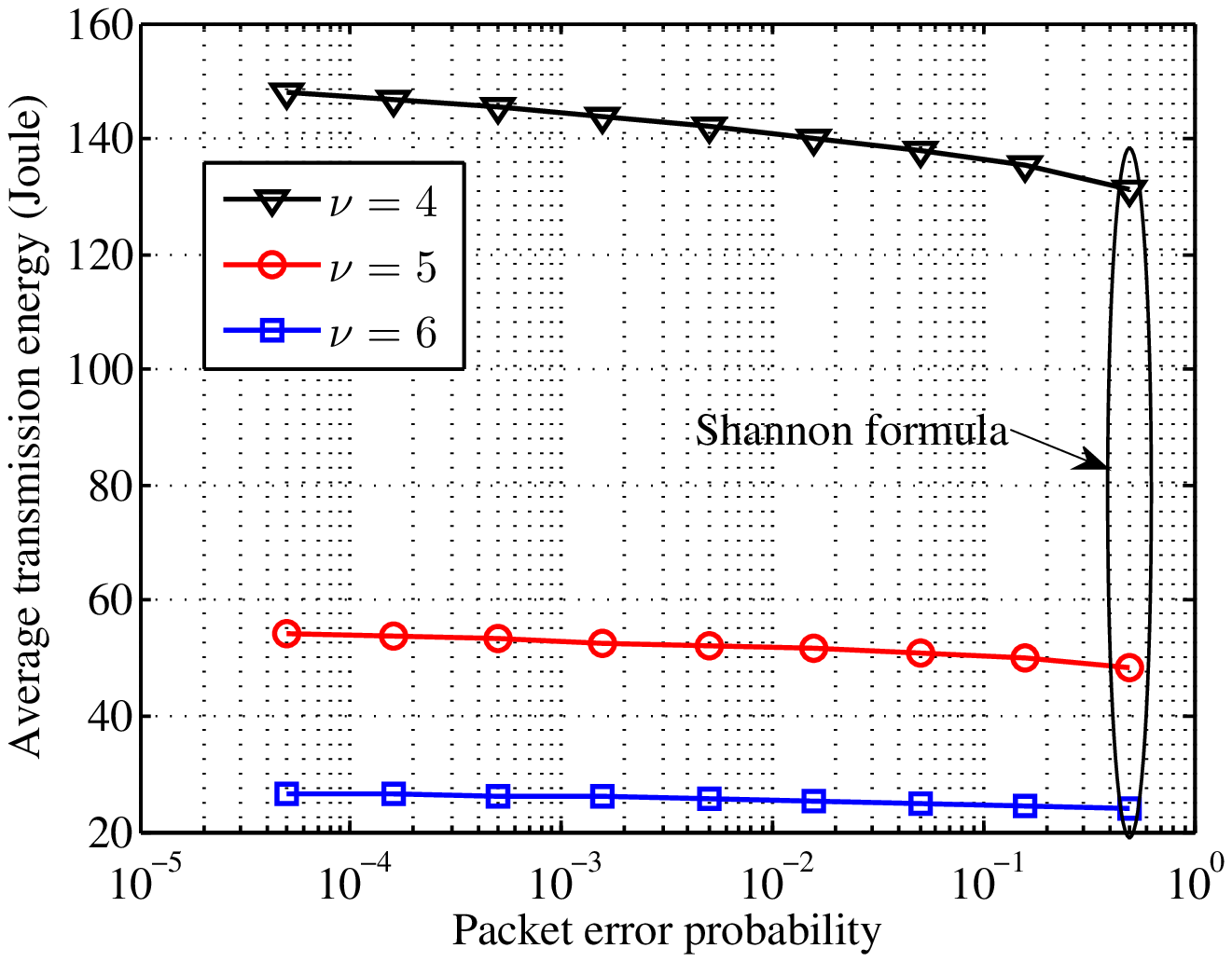}}
  \subfigure[]{
    \label{EU-Epsilon-n10} 
    \includegraphics[width=0.48\linewidth]{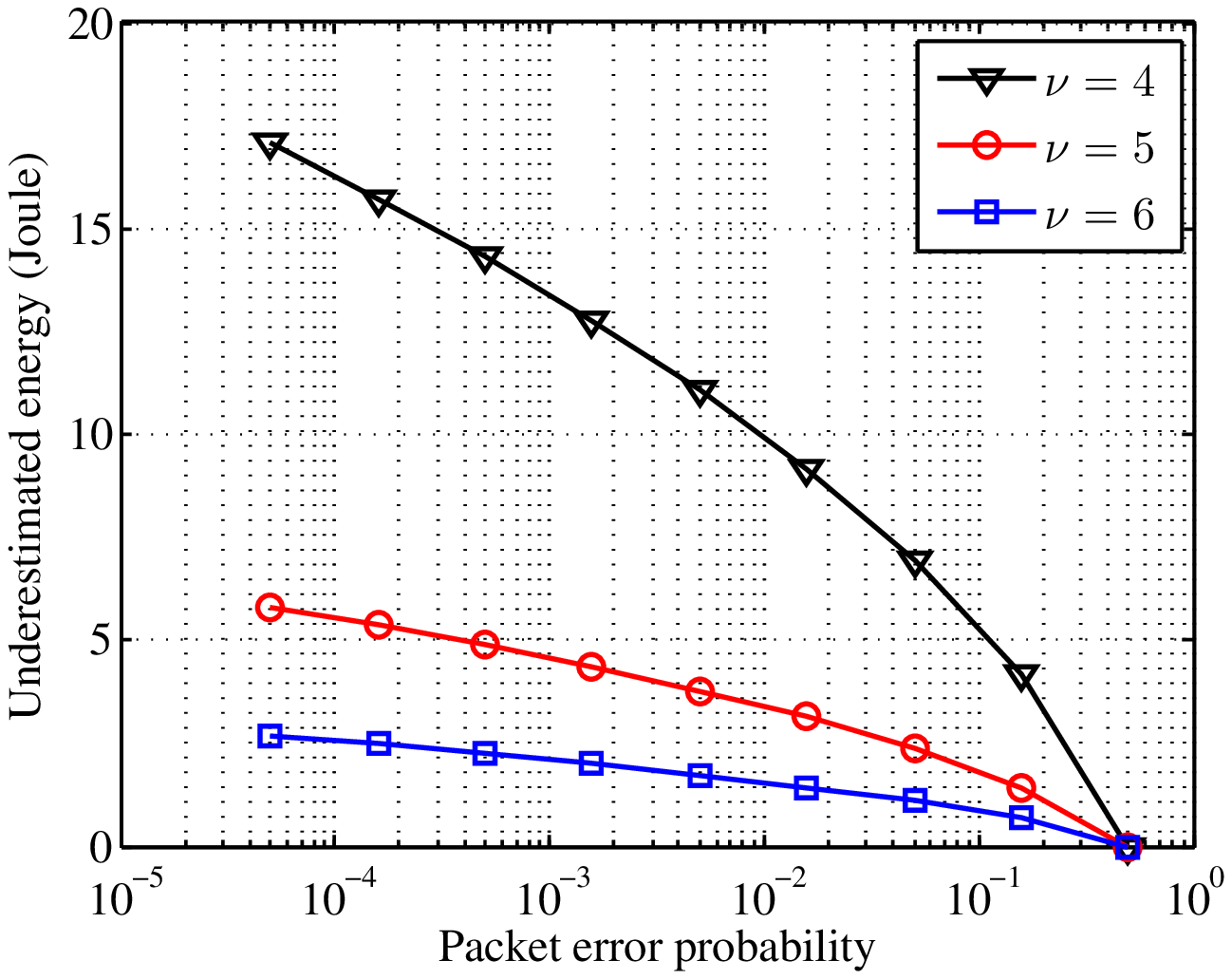}}
  \caption{Transmission energy and underestimated energy vs. packet error probability, for $n = 10$, $K =5$, $\sigma = 10$ and different values of $\nu$ (packet arrival interval).}
\end{figure}

Fig. \ref{EC-Epsilon-n10} and Fig. \ref{EU-Epsilon-n10}
respectively show the average transmission
energy and the corresponding underestimated energy versus packet
error probability $\epsilon$, under different values of $\nu$ for
controlling the packet arrival interval.
It can be seen from Fig. \ref{EC-Epsilon-n10} that a shorter
packet arrival interval results in more energy consumption.
This is because shorter packet arrival intervals enforces shorter code blocklength,
and more energy will be consumed by the decreasing property of the energy function.
Due to the same reason, a smaller value of $\nu$ results in more serious energy
underestimation as shown in Fig. \ref{EU-Epsilon-n10}.

\subsection{Performance Comparison of Online and Offline Algorithms}
Here we compare the performance of the proposed two offline algorithms and the online algorithm.
As a benchmark, a \emph{myopic} online scheme is also considered. In this myopic scheme,
whenever a data packet is ready to transmit, the scheduler always encodes the packet with a
maximum blocklength which is equal to the difference between the current time and the packet deadline.
We set $\nu = 4$, $n = 10$, $\epsilon = 5\times10^{-4}$ and $K=10$.
Under this setting, the energy function is strictly convex.
Therefore, both the MLWF algorithm and SUM method can be employed to solve this convex problem and obtain the optimal solution.
In the SUM method, we used the MLWF algorithm to handle the subproblem \eqref{PSUM}. 
We assume that there are three packets arrived before time 0.

\begin{figure}[!t]
  \centering
  \includegraphics[scale =0.7]{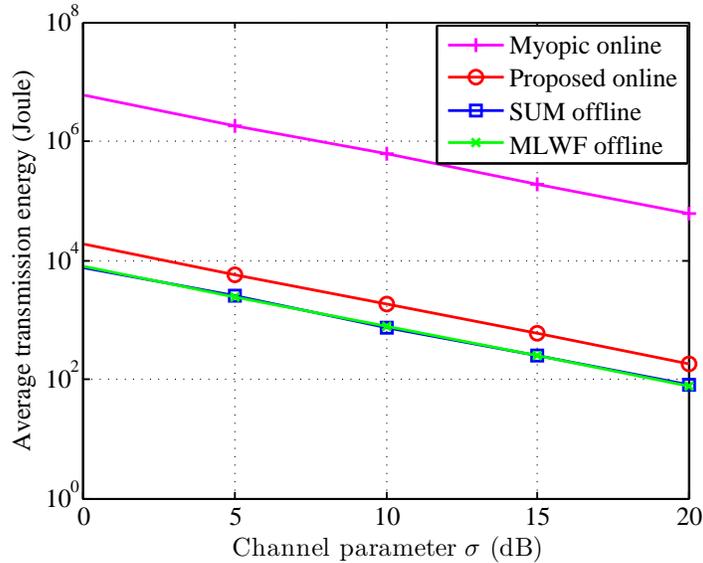}\\
  \caption{Energy consumptions for different scheduling algorithms when $\nu = 4$, $n = 10$, $\epsilon = 5\times10^{-4}$ and $K=10$}\label{EC-sigma-dB}
\end{figure}

Fig. \ref{EC-sigma-dB} presents the average transmission energies of different
scheduling algorithms under consideration.
As seen from the figure, the two proposed offline scheduling algorithms attain
the same average energy consumption and perform better than
the other two online algorithms. This is because the offline algorithms are
under the ideal assumptions that global knowledge of the packet arrivals is completely available.
Without global knowledge, one can see that the proposed
online algorithm still performs much better than the myopic algorithm.
The reason is that the proposed online algorithm always chooses a longest possible scheduling window
to fully utilize the deadline information of all arrived packets.
Finally, as observed from the figure without surprise, the average energy consumption decreases when the channel gain increases.

\section{Conclusions}\label{sec: conclusion}
In this paper, we have investigated the energy-efficient packet scheduling problem by considering the new finite-blocklength channel capacity formula.
While the newly formulated scheduling problem is inherently more complicated than the traditional designs that use the Shannon capacity formula, we have analytically shown that the packet transmission energy is a monotonically decreasing and convex function of the code blocklength as long as the code blocklength is properly upper bounded.
These appealing properties are therefore automatically valid for scenarios where the packets are subject to short delay constraints.
To solve the packet scheduling problem efficiently, we have also presented the MLWF algorithm and the SUM algorithm
for offline packet scheduling as well as a rolling-window based online algorithm for real-time packet scheduling.
The presented simulation results have shown that, in comparison with the proposed finite-blocklength packet scheduling design, the traditional design using the Shannon capacity formula can considerably underestimate the required transmission energy for reliable communications. We have also shown that the proposed online algorithm substantially outperforms the myopic scheduling schemes.

In the current work, we have assumed long or medium-range wireless communications where the energy consumption is mainly contributed by the data transmission power and the circuit power  due to
signal processing is negligible. However, for other scenarios (such as low-range
wireless communications), the circuit power may have to be taken into account in the packet scheduling design; see \cite{Wang-2015, Jin-2014}. Generalization of
the current work to that with circuit power consumption would be an interesting
direction for future research.

\begin{appendices}
\section{Proof of Proposition \ref{L1}}\label{app1}

For notational simplicity, we remove the subindex $k$ of all variables and let $x=ph>0$. Moreover, we write
$a \circeq b$ if $ab>0$ (i.e., $a$ and $b$ have the same sign).
From \eqref{P1-0}, we also define
\begin{align}\label{app1-2}
  \Fc(m,x) \triangleq   m\ln(x+1)  -\sqrt{m}\frac{\sqrt{x(x+2)}}{x+1}Q^{-1}(\epsilon) - N\ln 2 = 0.
\end{align}

{\bf Proof of Proposition \ref{L1}(a):}
Firstly, note that the left-hand side of \eqref{app1-2} is a quadratic equation of $\sqrt{m}$.
Let $\alpha = \ln(x+1)$, and $\beta = \frac{\sqrt{x(x+2)}}{x+1}Q^{-1}(\epsilon)$, the positive root of \eqref{app1-2} can be given by
\begin{subequations}\label{app1-3}
\begin{align}
  \sqrt{m} &= \frac{\beta+\sqrt{\beta^2+4N\alpha\ln 2}}{2\alpha}\label{app1-3-1}\\
           &> \frac{\beta}{\alpha} = \frac{\sqrt{x(x+2)}Q^{-1}(\epsilon)}{(x+1)\ln(x+1)},\label{app1-3-2}
\end{align}
\end{subequations}
where the inequality holds when $Q^{-1}(\epsilon)>0$, i.e., $\epsilon \in (0,0.5)$.
Secondly, according to the implicit function theorem \cite{Krantz_Parks02}, we have
\begin{align}\label{app1-4}
  \frac{\partial P(m)}{\partial m} \circeq \frac{\partial x }{\partial m} = -\frac{\frac{\partial \Fc}{\partial m}}{\frac{\partial \Fc}{\partial x}}.
\end{align}

Based on \eqref{app1-2}, it can be readily shown that
\begin{subequations}\label{app1-5}
\begin{align}%
   \Fc_m'\triangleq \frac{\partial \Fc}{\partial m}
         &= \ln(x+1) - \frac{Q^{-1}(\epsilon)}{2\sqrt{m}}\frac{\sqrt{x(x+2)}}{x+1}\label{app1-5-1}\\
         &= \frac{1}{2}\left(\ln(x+1) + \frac{N\ln 2}{m}\right) > 0,\label{app1-5-2}
\end{align}
\end{subequations}
and
\begin{subequations}\label{app1-6}
\begin{align}
    \Fc_x'\triangleq \frac{\partial \Fc}{\partial x}
         &= \frac{m}{x+1}- \frac{Q^{-1}(\epsilon)\sqrt{m}}{(x+1)^2\sqrt{x(x+2)}}\label{app1-6-1}\\
         &= \frac{\sqrt{m}}{x+1}\left(\sqrt{m}- \frac{Q^{-1}(\epsilon)}{(x+1)\sqrt{x(x+2)}}\right)\label{app1-6-2}\\
         &>  \frac{\sqrt{m}}{x+1}\left(\frac{\sqrt{x(x+2)}Q^{-1}(\epsilon)}{(x+1)\ln(x+1)} - \frac{Q^{-1}(\epsilon)}{(x+1)\sqrt{x(x+2)}}\right)\label{app1-6-3}\\
         &=  \frac{\sqrt{m}Q^{-1}(\epsilon)}{\sqrt{x(x+2)}(x+1)^2\ln(x+1)}\Big(x(x+2) -\ln(x+1)\Big) > 0, \label{app1-6-4}
\end{align}
\end{subequations}
where the inequality \eqref{app1-6-3} holds due to \eqref{app1-3-2} and one can easily check $x(x+2) -\ln(x+1)>0$ for $x>0$.

Thus, we can conclude that for any given $x>0$, $\epsilon\in(0,0.5)$ and $m\geq\hat{m}$, we always have $\frac{\partial x}{\partial m} < 0$ from \eqref{app1-4}, \eqref{app1-5-2} and \eqref{app1-6-4}.
That is equivalently to say, if \eqref{P1-0} holds true with $\epsilon\in(0,0.5)$, then $P(m)$ is always decreasing with $m$, which completes this proof. \hfill $\blacksquare$

{\bf Proof of Proposition \ref{L1}(b):}  To prove the monotonicity of the energy function $E(m)$, we have the derivative of $E(m)$ as
\begin{subequations}\label{app2-1}
\begin{align}
 &\frac{\partial E(m)}{\partial m}= \frac{\partial mP(m)}{\partial m} \label{app2-1-1}\\
 &\circeq \frac{\partial mx}{\partial m} = x + m\frac{\partial x}{\partial m} = x - m\frac{\Fc'_m}{\Fc'_x}\label{app2-1-2}\\
 &\circeq  x\Fc'_x - m\Fc'_m\label{app2-1-3}\\
 &=  \frac{x\sqrt{m}}{x+1} \left(\sqrt{m}- \frac{Q^{-1}(\epsilon)}{(x+1)\sqrt{x(x+2)}}\right)
 - m\left(\ln(x+1) - \frac{Q^{-1}(\epsilon)}{2\sqrt{m}}\frac{\sqrt{x(x+2)}}{x+1}\right) \label{app2-1-4}
 \\
 &= \left(\frac{x}{x+1} - \ln(x+1)\right)m + \frac{\sqrt{x(x+2)}}{x+1}\left( \frac{1}{2} - \frac{1}{(x+1)(x+2)}\right)Q^{-1}(\epsilon)\sqrt{m} \label{app2-1-5}\\
 &\circeq \left(\frac{x}{x+1} - \ln(x+1)\right) + \frac{\sqrt{x(x+2)}}{x+1}\left( \frac{1}{2} - \frac{1}{(x+1)(x+2)}\right)\frac{Q^{-1}(\epsilon)}{\sqrt{m}} \label{app2-1-6}
 \\
 &< \left(\frac{x}{x+1} - \ln(x+1)\right)  + \frac{\sqrt{x(x+2)}}{x+1}\left( \frac{1}{2} - \frac{1}{(x+1)(x+2)}\right)\frac{Q^{-1}(\epsilon)}{\sqrt{\hat{m}}} \label{app2-1-7}
 \\
 &< \left(\frac{x}{x+1} - \ln(x+1)\right)  + \left( \frac{1}{2} - \frac{1}{(x+1)(x+2)}\right)\frac{Q^{-1}(\epsilon)}{\sqrt{\hat{m}}} \label{app2-1-8}\\
 &< \left(\frac{x}{x+1} - \ln(x+1)\right)  + \frac{1}{2}\frac{Q^{-1}(\epsilon)}{\sqrt{\hat{m}}} \label{app2-1-9}
\end{align}
\end{subequations}
where \eqref{app2-1-3} and \eqref{app2-1-6} hold due to $\Fc'_x > 0$ (see \eqref{app1-6-4}) and $m>0$, respectively.
\eqref{app2-1-7} holds because of $m > \hat m$. In addition, \eqref{app2-1-8} and \eqref{app2-1-9} hold since $\frac{\sqrt{x(x+2)}}{x+1} = \frac{\sqrt{(x+1)^2 -1}}{x+1} < 1$ and $\frac{1}{2} > \frac{1}{2} - \frac{1}{(x+1)(x+2)} > 0$ for $x>0$, respectively.

From \eqref{app2-1-9}, to make $\frac{\partial E(m)}{\partial m} <0$, it is sufficient to have
\begin{align}\label{app2-2}
 \left(\frac{x}{x+1}-\ln(x+1)\right)  + \frac{\tau}{2} < 0,
\end{align}
where $\tau\triangleq \frac{Q^{-1}(\epsilon)}{\sqrt{\hat{m}}}$.
To find the range of $x$ satisfying \eqref{app2-2}, the following function is defined from the left-hand side of \eqref{app2-2} as
\begin{equation}\label{app2-3}
  g(x) \triangleq \frac{x}{x+1}-\ln(x+1) = 1-\frac{1}{x+1}-\ln(x+1),
\end{equation}
with $x\geq0$, $g(0) = 0$, and $g(\infty) = -\infty$.
From \eqref{app2-3}, we have the first-order derivative of $g(x)$
\begin{align}\label{app2-4}
g'(x) = \frac{1}{(x+1)^2}-\frac{1}{(x+1)} = -\frac{x}{(x+1)^2} < 0.
\end{align}
Thus, we can obtain that $g(x)$ is decreasing with $x$ for $x\geq0$ and there exists only one $x^{\ast}$ satisfying $g(x^{\ast})=-\frac{\tau}{2}$, which implies that $g(x)+\frac{\tau}{2}<0$ for $x>x^{\ast}$.
Mathematically, one can obtain $x^{\ast}$ by the following steps
\begin{subequations}\label{app2-5}
\begin{align}
 & 1-\frac{1}{x^{\ast}+1}-\ln(x^{\ast}+1) = -\frac{\tau}{2}\label{app2-5-1}\\
 \Longleftrightarrow & \exp\left(-\frac{1}{x^{\ast}+1}+\ln\left(\frac{1}{x^{\ast}+1}\right)\right) = \exp\left(-1-\frac{\tau}{2}\right)\label{app2-5-2}\\
 \Longleftrightarrow & \left(-\frac{1}{x^{\ast}+1}\right)\exp\left(-\frac{1}{x^{\ast}+1}\right) = -\exp\left(-1-\frac{\tau}{2}\right)\label{app2-5-3}\\
 \Longleftrightarrow & -\frac{1}{x^{\ast}+1} = \Wc\left(-\exp\left(-1-\frac{\tau}{2}\right)\right)\label{app2-5-4}\\
 \Longleftrightarrow &~ x^{\ast} = -\frac{1}{\Wc\left(-\exp(-1-\frac{\tau}{2})\right)} - 1,\label{app2-5-5}
\end{align}
\end{subequations}
where $\Wc(z)$ is the solution to $\Wc(z)\exp(\Wc(z)) = z$, i.e., the Lambert W function, and the symbol $\Longleftrightarrow$ means ``if and only if".

We therefore conclude that if $x = ph > -\frac{1}{\Wc\left(-\exp(-1-\frac{\tau}{2})\right)} - 1$, then $\frac{\partial E(m)}{\partial m} <0$ holds.
Moreover, since $P(m)$ is decreasing with $m$, the function $\Xc(m) = P(m)h$ in \eqref{Xc} is also decreasing with $m$.
Thus, we have if $\hat m\leq m \leq \Xc^{-1}\left(-\frac{1}{\Wc\left(-\exp(-1-\frac{\tau}{2})\right)} - 1\right)$, the energy function $E(m)$ is decreasing with $m$. \hfill $\blacksquare$

\section{Proof of Theorem \ref{T1}}\label{app3}
To prove the convexity of energy function $E(m)$, we have the second derivative of $E(m)$ as
\begin{align}\label{app3-1}
 \frac{\partial^{2} E(m)}{\partial m^{2}} \circeq \frac{\partial^{2} mx}{\partial m^{2}} = \frac{\partial (\frac{\partial mx}{\partial m})}{\partial m} = \frac{\partial (x + m\frac{\partial x}{\partial m}) }{\partial m}  = 2\frac{\partial x}{\partial m} + m \frac{\partial^{2} x}{\partial m^{2}}.
\end{align}
Now we apply the implicit function theorem. From \eqref{app1-4}, \eqref{app1-5-1} and \eqref{app1-6-1}, we have
\begin{equation}\label{app3-2}
  \frac{\partial x}{\partial m} = -\frac{\Fc'_{m}}{\Fc'_{x}},
\end{equation}
and
\begin{equation}\label{app3-3}
  \frac{\partial^{2} x}{\partial m^{2}} = -\frac{\partial}{\partial m}\left(\frac{\Fc'_{m}}{\Fc'_{x}}\right) = -\frac{\Fc''_{m}\Fc'_{x} - \Fc''_{x}\frac{\partial x}{\partial m}\Fc'_{m}}{(\Fc'_{x})^{2}} = -\frac{1}{\Fc'_{x}} \left(\Fc''_{m} + \Fc''_{x}\left(\frac{\Fc'_{m}}{\Fc'_{x}}\right)^{2}\right).
\end{equation}
Plugging \eqref{app3-2} and \eqref{app3-3} into \eqref{app3-1}, we have
\begin{subequations}\label{app3-4}
\begin{align}
  \frac{\partial^{2}mx}{\partial^{2}m} & = -2\frac{\Fc'_{m}}{\Fc'_{x}} -\frac{m}{\Fc'_{x}} \left(\Fc''_{m} + \Fc''_{x}\left(\frac{\Fc'_{m}}{\Fc'_{x}}\right)^{2}\right)\label{app3-4-1}\\
  & = \frac{1}{\Fc'_{x}} \left( -2\Fc'_{m} - m\Fc''_{m} - m \Fc''_{x}\left(\frac{\Fc'_{m}}{\Fc'_{x}}\right)^{2}\right)\label{app3-4-2}\\
  & \circeq  -2\Fc'_{m} - m\Fc''_{m} - m \Fc''_{x}\left(\frac{\Fc'_{m}}{\Fc'_{x}}\right)^{2}\label{app3-4-3}
\end{align}
\end{subequations}
where \eqref{app3-4-3} holds due to $\Fc'_{x} > 0$ in \eqref{app1-6}.

Next, we respectively analyze the lower bound of the terms in \eqref{app3-4-3}.
First, from \eqref{app1-5-2}, we obtain the lower bound of the first two terms in \eqref{app3-4-3}
\begin{subequations}\label{app3-6}
\begin{align}
  -2\Fc'_{m} - m\Fc''_{m} & = -\ln(1+x) - \frac{N\ln2}{m} - m\left(-\frac{N\ln2}{2m^{2}}\right) \label{app3-6-1}\\
  & = -\ln(1+x) - \frac{N\ln2}{2m} \label{app3-6-2}\\
  & = -\ln(1+x) - \frac{\ln(x+1)m  -\frac{\sqrt{x(x+2)}}{x+1}Q^{-1}(\epsilon)\sqrt{m} }{2m} \label{app3-6-3}\\
  & = -\frac{3}{2}\ln(1+x) + \frac{\sqrt{x(x+2)}}{2\sqrt{m} (x+1)}Q^{-1}(\epsilon) \label{app3-6-4}\\
  & > -\frac{3}{2}\ln(1+x), \label{app3-6-5}
\end{align}
\end{subequations}
where \eqref{app3-6-3} holds due to the equality \eqref{app1-2}, and \eqref{app3-6-5} holds due to $x>0$.
From \eqref{app3-6-2}, we can see that $-2\Fc'_{m} - m\Fc''_{m} < 0$.
Thus, to make $\frac{\partial^{2}mx}{\partial^{2}m} > 0$ in \eqref{app3-4-3}, we must have $- m \Fc''_{x}\left(\frac{\Fc'_{m}}{\Fc'_{x}}\right)^{2} > 0$.

To have $- m \Fc''_{x}\left(\frac{\Fc'_{m}}{\Fc'_{x}}\right)^{2} > 0$, let us analyze the bounds on $\Fc'_{m}$, $\Fc'_{x}$, and $\Fc''_{x}$, respectively.
First, from \eqref{app1-5-1}, we have
\begin{subequations}\label{app3-7}
\begin{align}
  \Fc'_{m} & = \ln(1+x) - \frac{\sqrt{x(x+2)}}{2\sqrt{m}(x+1)}Q^{-1}(\epsilon) \label{app3-7-1}\\
         & > \ln(1+x) - \frac{Q^{-1}(\epsilon)}{2\sqrt{m}} \label{app3-7-2}
\end{align}
\end{subequations}
where \eqref{app3-7-2} holds since $\frac{\sqrt{x(x+2)}}{x+1} = \frac{\sqrt{(x+1)^2 -1}}{x+1} < 1$ for $x>0$.
Second, from \eqref{app1-6-1}, we have
\begin{subequations}\label{app3-8}
\begin{align}
    \Fc'_{x} &= \frac{m}{x+1}- \frac{Q^{-1}(\epsilon)\sqrt{m}}{(x+1)^2\sqrt{x(x+2)}}\label{app3-8-1}\\
           & < \frac{m}{x+1}\label{app3-8-2}
\end{align}
\end{subequations}
Third, based on \eqref{app3-8-1}, we find 
\begin{subequations}\label{app3-9}
\begin{align}
  \Fc''_{x} &= -\frac{m}{(x + 1)^{2}} + \frac{Q^{-1}(\epsilon)\sqrt{m}}{(x+1)^4x(x+2)} \left(2(x+1)\sqrt{x(x+2)} + (x+1)^2\frac{2(x+1)}{2\sqrt{x(x+2)}}\right)\label{app3-9-1}\\
  &= -\frac{m}{(x + 1)^{2}} + \frac{Q^{-1}(\epsilon)\sqrt{m}}{(x+1)^3x(x+2)} \left( \frac{2x(x+2)+(x+1)^2}{\sqrt{x(x+2)}}\right)\label{app3-9-2}\\
  &= \frac{\sqrt{m}}{(x + 1)^{2}} \left( -\sqrt{m} + \frac{Q^{-1}(\epsilon)}{\sqrt{x(x + 2)}} \frac{3(x + 1)^{2}-2}{(x + 1)((x + 1)^{2}-1)} \right)\label{app3-9-3}
\end{align}
\end{subequations}
When $x \geq 1$, we have
\begin{align}\label{app3-10}
 \frac{3(x + 1)^{2}-2}{(x + 1)((x + 1)^{2}-1)} < \frac{3(x + 1)}{(x + 1)^{2}-1}
        \leq \frac{3(x + 1)}{(1 + 1)^{2}-1}
        = x + 1.
\end{align}
By plugging the inequality \eqref{app3-10} into \eqref{app3-9-3}, we have
\begin{subequations}\label{app3-11}
\begin{align}
  \Fc''_{x} &< \frac{\sqrt{m}}{(x + 1)^{2}} \left( -\sqrt{m} + \frac{x+1}{\sqrt{x(x + 2)}}Q^{-1}(\epsilon) \right)\label{app3-11-1}\\
  &< \frac{\sqrt{m}}{(x + 1)^{2}} \left( -\sqrt{m} + \sqrt{\frac{x+2}{x}} Q^{-1}(\epsilon) \right)<0\label{app3-11-2}
\end{align}
\end{subequations} for $x \geq 1$, where the last inequality is obtained by assuming $\sqrt{\frac{x+2}{x}} Q^{-1}(\epsilon) < \sqrt{\hat{m}}  < \sqrt{m}$, i.e, $x > \frac{2\tau^2}{1-\tau^2}$ with $\tau = \frac{Q^{-1}(\epsilon)}{\sqrt{\hat{m}}}$.
Note that $\Fc'_{x}$ in \eqref{app3-8-2} is positive from \eqref{app1-6}, and the right-hand side of \eqref{app3-7-2} is positive if $x > \exp(\frac{\tau}{2})-1$.
Then if $x > \max\left\{\frac{2\tau^2}{1-\tau^2}, \exp\left(\frac{\tau}{2}\right)-1, 1\right\}$, by combining \eqref{app3-7-2}, \eqref{app3-8-2} and \eqref{app3-11-2}, we obtain
\begin{subequations}\label{app3-12}
\begin{align}
- m \Fc''_{x}\left(\frac{\Fc'_{m}}{\Fc'_{x}}\right)^{2}
> & - m \frac{\sqrt{m}}{(x + 1)^{2}} \left( -\sqrt{m} + \sqrt{\frac{x + 2}{x}}Q^{-1}(\epsilon)\right)
   \left(\frac{x+1}{m}\left(\ln(1+x) - \frac{Q^{-1}(\epsilon)}{2\sqrt{m}}\right)\right)^2 \label{app3-12-1}\\
= & \left( 1- \sqrt{\frac{x + 2}{x}} \frac{Q^{-1}(\epsilon)}{\sqrt{m}}\right) \left(\ln(1+x) - \frac{Q^{-1}(\epsilon)}{2\sqrt{m}}\right)^2 \label{app3-12-2}\\
\geq & \left( 1- \sqrt{3} \frac{Q^{-1}(\epsilon)}{\sqrt{m}}\right) \left(\ln(1+x) - \frac{Q^{-1}(\epsilon)}{2\sqrt{m}}\right)^2 \label{app3-12-3}\\
> & \left( 1- \sqrt{3} \frac{Q^{-1}(\epsilon)}{\sqrt{\hat{m}}}\right) \left(\ln(1+x) - \frac{Q^{-1}(\epsilon)}{2\sqrt{\hat{m}}}\right)^2 \label{app3-12-4}\\
= & \left( 1- \sqrt{3} \tau\right) \left(\ln(1+x) - \frac{\tau}{2}\right)^2 \label{app3-12-5}
\end{align}
\end{subequations}
where \eqref{app3-12-3} and \eqref{app3-12-4} hold due to $\frac{x + 2}{x}\leq3$ for $x\geq1$ and $m > \hat{m}$, respectively. Note that \eqref{app3-12-5} is positive if $1- \sqrt{3} \tau > 0$.

Finally, by plugging \eqref{app3-6-5} and \eqref{app3-12-5} into \eqref{app3-4-3}, we have
\begin{subequations}\label{app3-13}
\begin{align}
  \frac{\partial^{2}E(m)}{\partial^{2}m} & \circeq -2\Fc'_{m} - m\Fc''_{m} - m \Fc''_{x}\left(\frac{\Fc'_{m}}{\Fc'_{x}}\right)^{2}\label{app3-13-1}\\
  & > -\frac{3}{2}\ln(1+x) + \left( 1- \sqrt{3} \tau\right) \left(\ln(1+x) - \frac{\tau}{2}\right)^2\label{app3-13-2}\\
  & = -\frac{3}{2}\left(\frac{\tau}{2} + y\right) + \left( 1- \sqrt{3} \tau\right) y^2\label{app3-13-3}
\end{align}
\end{subequations}
where $y \triangleq \ln(1+x) - \frac{\tau}{2}$.
From \eqref{app3-13-3}, to make $\frac{\partial^{2}E(m)}{\partial^{2}m} > 0$, it is sufficient to have
\begin{align}\label{app3-14}
 \left( 1- \sqrt{3} \tau\right) y^2 -\frac{3}{2}y - \frac{3\tau}{4} >0.
\end{align}
After solving \eqref{app3-14}, we obtain that, when $1- \sqrt{3} \tau > 0$, i.e., $0 <\tau < \frac{\sqrt{3}}{3}$, then
\begin{align}
 y > \eta(\tau) \triangleq \frac{3+\sqrt{9+12\tau(1-\sqrt{3}\tau)}}{4(1- \sqrt{3}\tau)}.
\end{align}

Thus, under the condition $0<\tau < \frac{\sqrt{3}}{3}$, if $x = ph > \max\{\exp(\eta(\tau) + \frac{\tau}{2})-1, \frac{2\tau^2}{1-\tau^2}, \exp\left(\frac{\tau}{2}\right)-1, 1\}$, then $\frac{\partial^{2}E(m)}{\partial^{2}m} > 0$.
On one hand, one can check that $\eta(\tau)$ and $\frac{2\tau^2}{1-\tau^2}$ are increasing with $\tau$ for $0<\tau < \frac{\sqrt{3}}{3}$, and $\exp(\eta(0))-1 >1 > \frac{2\tau^2}{1-\tau^2}$, thus we have $\max\{\exp(\eta(\tau) + \frac{\tau}{2})-1, \frac{2\tau^2}{1-\tau^2}, \exp\left(\frac{\tau}{2}\right)-1, 1\} = \exp(\eta(\tau) + \frac{\tau}{2})-1$ for $0<\tau < \frac{\sqrt{3}}{3}$.
In addition, since $\Xc(m) = P(m)h$ is decreasing with $m$ from Proposition \ref{L1}, we have that the energy function $E(m)$ is convex with $m$ when $\hat m \leq m \leq \Xc^{-1}\left(\exp\left(\eta(\tau) + \frac{\tau}{2}\right) -1\right)$.
\hfill $\blacksquare$

\end{appendices}
{
\small
\bibliographystyle{IEEEtran}
\bibliography{Ref_PTFBC}
}

\end{document}